\documentclass{article}
\usepackage{epsf,epsfig}
\usepackage{graphics,color}
\usepackage{amsmath}
\usepackage{amssymb}
\numberwithin{equation}{section}
\DeclareMathAlphabet{\mathpzc}{OT1}{pzc}{m}{it}
\DeclareMathAlphabet{\mathcalligra}{T1}{calligra}{m}{n}
\newenvironment{frcseries}{\fontfamily{frc}\selectfont}{}
\newcommand{\textfrc}[1]{{\frcseries#1}}
\newcommand{\Om}{\Omega}
\def\RR{{\mathbb{R}}}
\def\NN{{\mathbb{N}}}

\def\rot{\nabla\times}
\def\rott{\times\nabla}

\newcommand{\Lr}{\mathcal{L}}
\newcommand{\lr}{\textfrc{l}}
\newcommand{\OmLr}{\Omega_\mathcal{L}}

\newcommand{\strain}{\mathcal{E}^{\star}}
\newcommand{\strains}{\mathcal{E}^{\star s}}

\newcommand{\nl}{\newline}

\newtheorem{theorem}{Theorem}[section]{\bf}{\it}
{\bf}{\it}
\newtheorem{lemma}{Lemma}[section]{\bf}{\it}
\newtheorem{Def}{Definition}[section]{\bf}{\it}
\newtheorem{Ass}{Assumption}[section]{\bf}{\it}
\newtheorem{Not}{Notations}[section]{\bf}{\it}
\newtheorem{Rem}{Remark}[section]{\bf}{\it}
{\bf}{\it}
\numberwithin{equation}{section}
\begin{document}
\title{A distributional approach to the geometry of $2D$ dislocations at the mesoscale\\\begin{normalsize}\textit{Part B: The case of a countable family of dislocations}\end{normalsize}}
\author{Nicolas Van Goethem$^{1}$\& Fran\c{c}ois Dupret$^{2}$ \\
$^{1}$Centro de Matem\'{a}tica e Aplica\c{c}\~{o}es Fundamentais,\\
Universidade de Lisboa,\\
Av. Prof. Gama Pinto, 1649-003 Lisboa, Portugal\\
$^{2}$CESAME, Universit\'e catholique de Louvain,\\
4 av. G. Lema\^{\i}tre, 1348 Louvain-la-Neuve,\ Belgium}
\maketitle
\date{}

\begin{abstract}
This paper develops a geometrical model of dislocations and disclinations in single crystals at the mesoscopic scale. In the continuation of previous work the distribution theory is used to represent concentrated effects in the defect lines which in turn form the branching lines of the multiple-valued elastic displacement and rotation fields. Fundamental identities relating the incompatibility tensor to the dislocation and disclination densities are proved in the case of countably many parallel defect lines, under global $2D$ strain assumptions relying on the geometric measure theory. Our theory provides the appropriate objective internal variables and the required mathematical framework for a rigorous homogenization from mesoscopic to macroscopic scale.
\end{abstract}

\section{Introduction}
\label{intro}

\subsection{Preliminaries and principal hypotheses}
\label{prelim}
The present paper provides a mathematical theory of the geometry of crystal dislocations and disclinations in continuation of the work of Van Goethem \& Dupret (2009$a$) where the general context of this research is detailed and which will be referred to as Part A in the sequel. In summary, the objective of these investigations is to develop a rigorous mathematical framework for the treatment of line defects in single crystals at the mesoscopic scale. As explained in Part A, concentrated effects in the defect lines and their neighbourhood have to be considered at this scale and this requires to make use of the distribution theory (Schwartz 1957) to handle the related fields (dislocation and disclination densities, contortion, incompatibility, etc.) and their relationships. Moreover, in view of the incompatibility of the elastic strain tensor in the presence of line defects, the associated rotation and displacement are multiple-valued fields whose branching lines are precisely the defect lines. The combined treatment of distributions and multivalued functions was addressed in Part A, where our theory was applied to the case of a set of isolated parallel, moving or not, line defects under the hypothesis of a $2D$ elastic strain field.

In this second paper, the case of countably many parallel defect lines is investigated. Therefore, instead of analysing the regularity of the elastic strain near an assumed isolated defect line, a more general abstract approach is selected with a view to defining the appropriate functional space to validate the main theorem relating the strain incompatibility to the defect densities.

\begin{figure}[h!]
\begin{center}
\includegraphics[width=3cm,height=1.5cm]{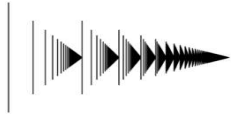}
\end{center} 
\caption{$\omega^2$ dislocation structure, where $\omega$ is the first ordinal transfinite number (Cantor 1915).\label{cantor}}
\end{figure}

Let us remark that our mesoscopic setting will be able to treat fine and complex dislocation structures since accumulation lines or points in the defective set will be allowed (such as typically the structures appearing in the work of Cantor (1915) on transfinite numbers, see figure \ref{cantor}). This feature represents a key ingredient of our theory since a tending to infinity number of defect lines unavoidably appears in the homogenization process from meso- to macro-scale. Moreover, when the defect lines exhibit a clustered mesoscopic structure, even if their actual number will always remain finite across any bounded area, it is much more convenient mathematically to consider a model where this structure may be infinitely refined.

Before any further development, precise definitions and assumptions are required. In general, the functional spaces used consist of distributions, Radon measures (Ambrosio et al. 2000, Mattila 1995), functions, etc., which can be considered as continuous functionals over a set of test-functions whose regularity determines the functional space properties. However particular care has to be given to avoiding undesirable  boundary effects.

\begin{Def}[functional spaces used]\label{fctspaces}
The crystal domain is an open connected set $\Om$, on which some mathematical elements (distributions, Radon measures, locally summable functions, etc.) are defined as linear functionals over an associated set of test-functions whose support is a compact subset of $\Om$. Henceforth, the qualification ``on $\Om$'' for these elements will mean in addition that extensions of these elements exist as functionals over all the test-functions having the desired properties and whose support is a compact subset of $\RR^3$.
\end{Def}

In other words, a distribution, Radon measure, locally summable function on $\Om$ will always be constrained to also be the restriction to $\Om$ of a mathematical element of the same type defined on the whole $\RR^3$. A simple $1D$ example can illustate this constraint. The function
\begin{equation}
 f:\RR^+_0\longrightarrow\RR,\quad x\longmapsto f(x)=1/x\ (x>0)\label{f}
\end{equation} 
is locally summable on $\RR^+_0$ in the classical sense but cannot be extended as a locally summable function over $\RR$, this resulting from its behaviour for $x\to 0^+$. According to definition \ref{fctspaces}, in the present paper this function will not be called locally summable on $\RR^+_0$. Indeed, in a context where locally summable functions are needed and where the physical domain of interest is $\RR^+_0$, no shift to the right of $f$ as defined by (\ref{f}) can provide a locally summable function on $\RR^+_0$ (with $g(x)=f(x-a), \ x>a>0$) whatever the definition of $g$ is for $0<x\leq a$. So, if to be acceptable the locally summable functions on $\RR^+_0$ are requested to exhibit the same properties near the origin as in the vicinity of their interior points (to avoid peculiar boundary effects for $x\to 0^+$), freely translating these functions to move the origin onto an interior point should be allowed, and hence $f$ cannot be accepted as locally summable on $\RR^+_0$ if $g$ cannot. This justifies definition \ref{fctspaces}.

Similarly, the same function $f$ defined by (\ref{f}) can be considered as the density of a Radon measure on $\RR^+_0$ in the classical sense (as acting against continuous test-functions whose support is compact and contained in $\RR^+_0$). However, again no shift to the right of $f$ can provide a Radon measure on $\RR^+_0$ since no extension to $\RR$ of $f$ as a Radon measure in the classical sense exists. So, in our work $f$ will not be considered as a Radon measure on $\RR^+_0$ according to definition \ref{fctspaces}.

Considering now $f$ as a distribution  on $\RR^+_0$ (as acting against $\mathcal{C}^\infty$ test-functions whose support is a compact subset of $\RR^+_0$), $f$ can be prolonged as a distribution on $\RR$ by defining for example the prolonged $f$ as the pseudo-function $Pf. (1/x)$, which is the distributional derivative of the function $F(x):=\log|x|$ and acts again a test-function by taking the Hadamard finite part of the resulting diverging integral. So, here $f$ can be considered as a distribution on $\RR^+_0$ in the sense of definition \ref{fctspaces}.

It should be noted that the spaces generated from definition \ref{fctspaces} are non-closed subspaces of classical spaces (the usual distributions, Radon measures, ... on $\Om$) and hence cannot share all their properties (completeness, etc.).
\nl

In the crystal domain $\Om$, the meso-scale physics will then be represented by a nowhere dense set of defect lines which in $2D$ are parallel to each other.
\begin{Def}[$2D$ mesoscopic defect lines]\label{lines2D}
At the meso-scale, a $2D$ set of dislocations and/or disclinations $\Lr\subset\Om$ is a closed set of $\Om$ (this meaning the intersection with $\Om$ of a closed set of $\RR^3$) formed by a countable union of parallel lines $L^{(i)}, i\in\mathcal{I}\subset\NN$, whose adherence is itself a countable union of lines  and where the linear elastic strain is singular. In the sequel, these lines will be assumed as parallel to the $z$-axis. 
\end{Def}

The present mesoscopic theory will be completely developed from the sole linear strain -- which itself could be defined from the stress field (although the stress-strain relationship is not used in the sequel) and therefore is an objective internal field.

\begin{Ass}[$2D$ mesoscopic elastic strain]\label{asstrain}
The linear strain $\mathcal{E}^\star_{mn}$ is a given symmetric $L^1_{loc}(\Om)$ tensor (in the sense of definition \ref{fctspaces}) prolonged by $0$ on the dislocation set $\Lr$, and such that $\partial_z\mathcal{E}^\star_{mn}=0$. Moreover, $\mathcal{E}^\star_{mn}$ is assumed as compatible on $\OmLr:=\Om\setminus\Lr$ in the sense that the incompatibility tensor defined by
\begin{eqnarray}
&&\hspace{-150pt}\mbox{\scriptsize{INCOMPATIBILITY:}}\hspace{45pt}\eta^\star_{kl}:=\epsilon_{kpm}\epsilon_{lqn}\partial_p\partial_q\mathcal{E}^\star_{mn},\label{eta}
\end{eqnarray}
where derivation is intended in the distribution sense, vanishes everywhere on $\OmLr$. Equivalently, in $2D$ there are real numbers $K, a_\alpha$ and $b$ such that the following equalities hold on $\OmLr$:
\begin{eqnarray}
\left\{\begin{array}{lll}\epsilon_{\alpha\gamma}\epsilon_{\beta\delta}\partial_\alpha\partial_\beta\mathcal{E}_{\gamma\delta}^\star=0, \\
\epsilon_{\alpha\beta}\partial_\alpha\mathcal{E}_{\beta z}^\star=K, \\ \mathcal{E}_{zz}^\star=a_\alpha x_\alpha+b.\end{array}\right.\label{eqcomplane}
\end{eqnarray}
\end{Ass}

In Part A,  key tensor fields were derived from the mesoscopic elastic strain field as order-$1$ distributions (acting on $\mathcal{C}^1_c$ test-functions with compact support).
\begin{Def}[Frank and Burgers tensors]\label{FBtensors}
\begin{eqnarray}
&&\hspace{-108pt}\mbox{\scriptsize{FRANK TENSOR:}}\hspace{45pt}\overline\partial_m\omega_k^\star:=\epsilon_{kpq}\partial_p\mathcal{E}_{qm}^\star,\label{delta_m_a}\\
&&\hspace{-108pt}\mbox{\scriptsize{BURGERS TENSOR:}}\hspace{41pt}\overline\partial_l b^\star_k:=\mathcal{E}^\star_{kl}+\epsilon_{kpq}(x_p-x_{0p})\overline\partial_l\omega^\star_q.\label{delta_lb_i}
\end{eqnarray}
\end{Def}
Line integration of the Frank and Burgers tensors in $\OmLr$ (i.e., outside the defect set) provides the index-$1$ multivalued rotation and Burgers vector fields $\omega^\star_k$ and $b^\star_k$ (with ``index-$1$'' meaning that their first derivatives in $\OmLr$, denoted by $\partial^{(s)}_j\omega^\star_k$ and $\partial^{(s)}_jb^\star_k$, are single-valued). These properties are summarized in the following theorem, whose proof is classical.
\begin{theorem}[multiple-valued displacement field]\label{displ}
From a symmetric smooth linear strain $\mathcal{E}^{\star}_{ij}$ on $\Om_\Lr$ and a point $x_0$ where displacement and rotation are given, a multivalued displacement field $u^\star_i$ of index $2$ (whose second derivatives are single-valued) can be constructed on $\Om_\Lr$ such that the symmetric part of the distortion $\partial^{(s)}_ju^\star_i$ is the single-valued strain tensor $\mathcal{E}^\star_{ij}$ while its skew-symmetric part is the multivalued rotation tensor
 $\omega^\star_{ij}:=-\epsilon_{ijk}\omega^\star_k$. Moreover, inside $\OmLr$ the gradient $\partial^{(s)}_j$ of the rotation and Burgers fields $\omega^\star_k$ and $b^\star_k=u^\star_k-\epsilon_{klm}\omega^\star_l(x_m-x_{0m})$ coincides with the Frank and Burgers tensors.
\end{theorem}
From this result, the Frank and Burgers vectors can be defined as invariants of any isolated defect line $L^{(i)}$ of $\Lr$.
\begin{Def}[Frank and Burgers vectors]\label{Burgers}
The Frank vector of the isolated defect line $L^{(i)}$ is the invariant
\begin{eqnarray}
\Om^{\star(i)}_k:=[\omega^\star_k]^{(i)},\label{frank}
\end{eqnarray}
while its Burgers vector is the invariant
\begin{eqnarray}
B^{\star(i)}_k:=[b^\star_k]^{(i)}=[u_k^\star]^{(i)}(x)-\epsilon_{klm}\Om^{\star(i)}_l(x_m-x_{0m}),\label{burgers}
\end{eqnarray}
with $[\omega^\star_k]^{(i)},[b^\star_k]^{(i)}$ and $[u_k^\star]^{(i)}$ denoting the jumps of $\omega^\star_k,b^\star_k$ and $u_k^\star$ around $L^{(i)}$.
\end{Def}

The case of non-isolated defect-lines represents a major issue of this work and will be resolved at a later stage. Besides their relationship with the multivalued rotation, Burgers and displacement fields, the Frank and Burgers tensors can be directly related to the strain incompatibility by use of (\ref{eta}), (\ref{delta_m_a}) and (\ref{delta_lb_i}).
\begin{theorem}
The distributional curls of the Frank and Burgers tensors are
\begin{eqnarray}
\epsilon_{ilj}\partial_l\overline\partial_j\omega^\star_k&=&\eta^\star_{ik},\label{sup1}\\
\epsilon_{ilj}\partial_l\overline\partial_jb^\star_k&=&\epsilon_{kpq}(x_p-x_{0p})\eta^\star_{iq},\label{sup2}
\end{eqnarray} 
with $\eta^\star_{ik}$ the incompatibility tensor.
\end{theorem}

From this theorem it results that single-valued rotation and Burgers fields $\omega^\star_k$ and $b^\star_k$ can be integrated on $\Om$ if, and only if the incompatibility tensor vanishes.
\nl

The dislocation and disclination densities are the basic physical tools used to model defect density at the meso-scale. In $2D$ (here meaning that $\partial_z\mathcal{E}^\star_{mn}=0$) the defect lines are parallel to the $z$-axis and hence only some components of the defect densities do not vanish.

\begin{Def}[$2D$ defect densities \footnote{Various notations are used in the literature to represent the defect densities. In particular, Nye (1953), Kr\"oner (1980) and Kleinert (1989) give different definitions of the dislocation density and contortion tensors (without considering disclinations in the first two cases). We here follow Kr\"oner's and Kleinert's notations for $\alpha^\star_{ij}$ and Nye's original definition of $\kappa^\star_{ij}$, with Nye's $\alpha^\star_{ij}$ here denoted by $\alpha^\star_{ji}$. It should be recalled that the term "contortion" was introduced by Kondo (1952).}]
\begin{eqnarray}
&&\hspace{-58pt}\mbox{\scriptsize{DISCLINATION DENSITY:}}\hspace{45pt}\Theta^\star_z:=\sum_{i\in\mathcal{I}}\Om^{\star (i)}_z\delta_{L^{(i)}},\label{disclindens1}\\
&&\hspace{-58pt}\mbox{\scriptsize{DISLOCATION DENSITY:}}\hspace{48pt}\Lambda^\star_k:=\displaystyle\sum_{i\in\mathcal{I}}B^{\star (i)}_k\delta_{L^{(i)}}  \quad(k=1\cdots 3).\label{dislocdens2}
\end{eqnarray}
\end{Def}
In general, the complete defect density tensors are denoted by $\Theta^\star_{ij}$ and $\Lambda^\star_{ij}$ with the indices $i$ and $j$ referring to the local defect line direction and Frank or Burgers vector, respectively. So in $2D$,
\begin{eqnarray}
\Theta^\star_{k\kappa}&=&\Theta^\star_{\kappa k}=0,\quad\Theta^\star_{zz}=\Theta^\star_z \label{disclindens1a}\\
\Lambda^\star_{\kappa k}&=&0,\hspace{41pt}\Lambda^\star_{zk}=\Lambda^\star_k.\label{dislocdens2a}
\end{eqnarray}
with $k=1\cdots 3$ and $\kappa=1,2$. The vanishing of $\Theta^\star_{k\kappa}$ was shown in Part A.
\nl

An additional important defect density tensor called the contortion was introduced by Nye (1953) from the work of Kondo (1952).
\begin{Def}[$2D$ mesoscopic contortion]
\begin{eqnarray}
&&\hspace{-58pt}\mbox{\scriptsize{CONTORTION:}}\hspace{87pt}\kappa_{ij}^\star:=\delta_{iz}\alpha^\star_j-\frac{1}{2}\alpha^\star_z\delta_{ij}\quad(i,j=1\cdots 3),\label{KR2} 
\end{eqnarray} 
where
\begin{eqnarray}
&&\hspace{68pt}\alpha^\star_k:=\Lambda^\star_k-\delta_{k\alpha}\epsilon_{\alpha\beta}\Theta^\star_z(x_\beta-x_{0\beta}).\label{KR3}
\end{eqnarray} 
\end{Def}

Here, $\alpha^\star_k$ is an auxiliary defect density measure associated with a general $3D$ tensor $\alpha^\star_{ij}=\Lambda_{ij}^\star+\epsilon_{jlm}\Theta^\star_{il}(x_m-x_{0m})$ such that in $2D$
\begin{eqnarray}
\alpha^\star_{\kappa k}&=&0,\hspace{41pt}\alpha^\star_{zk}=\alpha^\star_k,\label{dislocdens2b}
\end{eqnarray}
with $k=1\cdots 3$ and $\kappa=1,2$. The general $3D$ contortion is $\kappa^\star_{ij}=\alpha_{ij}^\star-\frac{1}{2}\alpha^\star_{mm}\delta_{ij}$, in such a way that in $2D$, $\kappa^\star_{\alpha j}$ vanishes if $\alpha\neq j$. 

\subsection{Objective of this work}
\label{sec:obj}
The principal objective of this work is to identify key distributional fields at the mesoscopic scale and to demonstrate their relationship in a rigorous functional analysis context, with a further view to providing the required framework for the homogenization of these fields and their relations to the macroscopic scale.

In \S\ref{distrset}, the main theorem of Part A (expressing the elastic strain incompatibility in terms of the defect densities and their gradients) will be extended to the case of a countable ensemble of parallel defect lines. To this end, besides the strain assumption \ref{asstrain} an additional assumption is made on the Frank tensor (\ref{delta_m_a}).

\begin{Ass}[mesoscopic strain assumption]\label{assfranktens}
The $2D$ strain $\mathcal{E}_{mn}^\star$ belongs to $L^1_{loc}(\Om)$ (in the sense of definition \ref{fctspaces}) and is compatible on $\OmLr$. Moreover, the $(m,z)$ components of the Frank tensor $\overline\partial_m\omega^\star_z=\epsilon_{\alpha\beta}\partial_\alpha\strain_{\beta m}\ (1\leq m\leq 3)$ are Radon measures on $\Om$, whose singular parts with respect to the Lebsgue measure are purely concentrated on $\Lr$ while their absolutely continuous parts have a $2D$ curl which itself is a Radon measure on $\Om$.
\end{Ass}

\begin{Rem}
No similar assumption could be made on the complete Frank tensor without contradicting the edge and screw dislocation examples of Part A. Moreover the absolutely continuous part of $\overline\partial_m\omega^\star_z$ cannot be required to be of bounded variation without contradicting the wedge disclination example of Part A. On the other hand, it will be seen that the sharp assumption \ref{assfranktens} is required to establish our theory in the general case of countably many dislocations.
\end{Rem}

Among several equivalent formulations, our main theorem then takes the following form.
\nl\nl
\textbf{Main theorem:}
\nl
\textbf{incompatibility decomposition for a countable set of $2D$ defect lines.}
\nl\nl
\textit{Under assumptions \ref{asstrain} and \ref{assfranktens}, incompatibility as defined by (\ref{eta}) is the vectorial first order distribution
\begin{eqnarray}
\eta_{ik}^\star=\eta_{ki}^\star=\Theta_{ik}^\star+\epsilon_{ilj}\partial_l\kappa^\star_{kj}.\label{etak}
\end{eqnarray}}

Let us also introduce here the main intermediate proposition needed for the proof of the above representation theorem. Under assumptions \ref{asstrain} and \ref{assfranktens}, the following decomposition theorem will be proved in the $2D$ case.
\begin{description}
\item[Theorem: $2D$ strain decomposition.]
\end{description}
\textit{Let the $2D$ strain tensor $\mathcal{E}^\star_{mn}$ be a compatible $L^1_{loc}$ tensor on $\OmLr$. Then the following decomposition holds:
\begin{eqnarray}
 \mathcal{E}^\star_{mn}=e^\star_{mn}+E^\star_{mn},\label{strain_decompa}
\end{eqnarray} 
where $e^\star_{mn}$ is everywhere compatible, whereas $E^\star_{mn}$ is a sum,
\begin{equation}
E^\star_{mn}=\sum_{i\in\mathcal{I}}E^{\star (i)}_{mn},\label{strain_sommea}
\end{equation} 
where each $E^{\star (i)}_{mn}\ (i\in\mathcal{I})$ is analytically known, compatible and smooth on $\Om\setminus L^{(i)}$, while $E^\star_{mn}$ is singular on $\Lr$ and compatible and smooth on $\OmLr$.}
\nl

From our main theorem, \S\ref{multi_inc} will then be devoted to introducing new mesoscopic distributional fields, namely the completed Frank and Burgers tensors, which will represent the appropriate objective internal variables after homogenization to the macro-scale, and to reformulate the main theorem in their terms. Conclusions will be drawn in \S\ref{concl}.

\section{Distributional analysis of incompatibility for a countable set of parallel dislocations}\label{distrset}
To capture the macro-scale physics, homogenization must be performed on a set of dislocation lines whose number tends to infinity in order to define diffuse defect density tensors. Therefore, assumption \ref{assfranktens} was introduced in a functional formulation that can be extended in some way from a set of defect lines (at the mesoscopic level) to a diffuse defect density (at the macroscopic level). 

The extension of our theory to a countable number of defect lines poses several technical problems. A first difficulty arises from the different kinds of convergence that could be required. Typically, considering a series of Dirac masses on $\lr_0=\mathcal{L}\cap\{z=z_0\}$, then its convergence as a measure implies that the sum of the weights must converge absolutely, but this is no longer the case if a (coarser) distributional convergence is required. A second example is provided by those distributions that are the gradient of a summable function. If these distributions are concentrated on isolated points, they must be the sum of Dirac masses, whereas this property might fail on a countable set. More generally, it is known (Schwartz 1957) that a concentrated first-order distribution on isolated points is a sum of weighted Dirac masses and Dirac mass derivatives, while a concentrated measure on a countable set is a sum of weighted Dirac masses. However, it is false to claim that a concentrated first-order distribution on a countable set is a sum of Dirac masses and Dirac mass derivatives, as $1D$ counter-examples can show. In general, a more complex mathematics governs the accumulation points of $\textfrc{l}_0$, and appropriate tools are required to extend the representation theorems of Part A to a countable set of defects.

\subsection{General strain decomposition property}
\label{sec:straindecomp}
In general any vector field can be decomposed into a solenoidal and an irrotational part, and this property can be easily extended to distributional fields. In this paper, the similar decomposition of any symmetric tensor field into a compatible and a solenoidal part will be used to extend the main theorem of Part A from isolated to countable dislocations \footnote{Kr\"{o}ner (1980) first observed that this decomposition provides a link between the dislocation density in a medium and the associated strain tensor incompatibility.}. In what follows, we will first give a proof of the decomposition existence in the general distributional case and then investigate its regularity in the $2D$ case. The main theorem will  be extended in a further section.

\begin{theorem}[standard decomposition of a symmetric tensor]
Any symmetric $2^{nd}$-order distribution tensor $E$ (or $E_{ij}$) can be decomposed into a compatible and a solenoidal symmetric part:
\begin{eqnarray}
E=E^c+E^s,\label{decomp1}
\end{eqnarray}
with
\begin{eqnarray}
\rot E^c \rott=0\quad\mbox{(compatible $E^c$)},\label{decomp2}
\end{eqnarray} 
and 
\begin{equation}
\nabla\cdot E^s=0\quad\mbox{(solenoidal $E^s$)}.\label{decomp3}
\end{equation} 
\end{theorem}
{\bf Proof.} $\blacktriangle$ Any tensor $E^s$ defined by the relation
\begin{equation}
E^s=\nabla\times F\times\nabla\label{decomp4}
\end{equation} 
is symmetric and solenoidal if $F$ is a symmetric tensor distribution. Then the reminder $E^c=E-E^s$ is compatible provided, after some calculations, $F$ satisfies the relation
\begin{eqnarray}
 \Delta\Delta F_{ij}+\partial_i\partial_j\partial_k\partial_l F_{kl}-\Delta\left(\partial_j\partial_k F_{ik}\right)-\Delta\left(\partial_i\partial_k F_{jk}\right)=\epsilon_{ikl}\epsilon_{jmn}\partial_k\partial_m E_{ln},\label{decomp5}
\end{eqnarray}
with $\Delta$ the Laplacian operator ($\Delta=\partial_i\partial_i$). If in addition the gauge condition
\begin{equation}
\nabla\cdot F=0\label{decomp6}
\end{equation} 
is imposed, then (\ref{decomp4}) reduces to the elliptic equation
\begin{equation}
\Delta\Delta F=\nabla\times E\times\nabla.\label{decomp7}
\end{equation} 
$\blacktriangle$  Therefore, to find the searched decomposition (\ref{decomp1}), (\ref{decomp2}), (\ref{decomp3}), it is sufficient to solve (\ref{decomp7}) for $F$ with the gauge condition (\ref{decomp6}). If $E$ is sufficiently regular, $F$ will simply be found by solving (\ref{decomp7}) with, among others, the $6$ boundary conditions $\nabla\cdot F=0$ and $\partial\left(\nabla\cdot F\right)/\partial n=0$. As a matter of fact, a solution exists because the operator $\Delta\Delta$ is elliptic, and this solution is divergence-free because taking the divergence of (\ref{decomp7}) provides the relation $\Delta\Delta\left(\nabla\cdot F\right)=0$ which, together with the boundary conditions implies that $\nabla\cdot F$ itself vanishes.

If $E$ is not sufficiently regular, $E$ can be approximated as a distribution by a family of $\mathcal{C}^\infty$ functions $E_\epsilon (\epsilon>0)$ with $E_\epsilon\to E$ for $\epsilon\to 0^+$ (Schwartz 1957). The family of equations obtained by replacing $E$  by $E_\epsilon$ in (\ref{decomp6}), (\ref{decomp7}) provides a family of solutions $F_\epsilon$ which tends to a suitable $F$ when $\epsilon\to 0^+$. {\hfill $\square$}

\subsection{First representation theorem of a $2D$ incompatible strain}
\label{sec:straindecomp1}

The previous section has shown that a distributional decomposition of the symmetric strain ${\mathcal{E}^\star}\in L^1_{loc}(\Om)$ into compatible and solenoidal distributional parts ${\mathcal{E}^\star}^c$ and ${\mathcal{E}^\star}^s$ always exists, with the right-hand side of (\ref{decomp7}) showing to be the incompatibility tensor. However, more regular solutions exist in the $2D$ case. Before proving them, the following result will be needed.

\begin{lemma}\label{L1} Let  $\delta^{(i)}$ stand for the Dirac measure at $\hat x^{(i)}\in\textfrc{l}_0$ and  $\displaystyle\sum_{i\in\mathcal{I}}C^{(i)}\delta^{(i)}$ be a Radon measure on $\Om_{z_0}=\Om\setminus\{z=z_0\}$ in the sense of definition \ref{fctspaces}. Then the sum of the weights $C^{(i)}$ is locally absolutely convergent, this meaning its absolute convergence on any bounded subset $\{\hat x^{(i)}, i\in\mathcal{I}'\}$ of $\lr_0$.
\end{lemma}
{\bf Proof.} Since $\displaystyle\sum_{i\in\mathcal{I}}C^{(i)}\delta^{(i)}$ is a Radon measure, then $\displaystyle\sum_{i\in\mathcal{I}'}C^{(i)}\delta^{(i)}$ is a finite Radon measure and the sum can be indifferently carried out on every permutation of $\mathcal{I}'$. Hence, taking a test-function wich equals $1$ on $\textfrc{l}_0$, the sum of the weights converges for every permutation of $\mathcal{I}'$ and is absolutely convergent.
{\hfill $\square$}

\begin{Rem}If $\displaystyle\sum_{i\in\mathcal{I}}C^{(i)}\delta^{(i)}$ were assumed to be a general distribution instead of a Radon measure, no such statement on the absolute convergence of the sum of the weights could be proved as the following simple $1D$ counter-example shows: selecting $\hat x^{(i)}=1/i\ (i\in \mathcal{I}=\NN_0)$ and $C^{(i)}=(-1)^{i+1}(1/i+1/(1+i))$ provides a distributionally convergent series $\displaystyle\sum_{i\in\mathcal{I}}C^{(i)}\delta^{(i)}$, since it is the derivative of the $L^1_{loc}$ converging series $1-\displaystyle\sum_{i\in\mathcal{I}}C^{(i)}(1-H^{(i)})$ with $H^{(i)}=H(x-\hat x^{(i)})$ and $H$ the step function, whereas the sum $\displaystyle\sum_{i\in\mathcal{I}}|C^{(i)}|$ does not converge.
\end{Rem}

\begin{Not}
Henceforth $\{\hat x^{(i)}, \ i\in\mathcal{I}\}$ will denote the set of points defining $\lr_0$.
\end{Not}

\begin{theorem}[regularity of the strain decomposition]\label{straindecomp}
 Let the strain and the Frank tensor satisfy assumptions  \ref{asstrain} and \ref{assfranktens}, and the dislocation set be defined according to definition \ref{lines2D}. Then the following decomposition holds:
\begin{eqnarray}
 \mathcal{E}^\star_{mn}= \mathcal{E}^{\star c}_{mn}+\mathcal{E}^{\star s}_{mn},\label{strain_decomp}
\end{eqnarray} 
where $\mathcal{E}^{\star c}_{mn}\in L^1_{loc}(\Om)$ is compatible, whereas $\mathcal{E}^{\star s}_{mn}\in L^1_{loc}(\Om)$ is solenoidal.
\end{theorem}
{\bf Proof.} Consider any $2D$ cut $\Om_{z_0}$ of $\Om$ and assume first that $\Om_{z_0}$ is bounded (extension to unbounded sets is direct). Since the strain is independent of $z$, in $2D$ it suffices to solve (\ref{decomp7}) and (\ref{decomp6}) for $F$ on $\Om_{z_0}$ with $E_{ij}=\strain_{ij}$. This will be achieved by solving an associated problem on $\Om_{z_0}$ by means of complex (but not necessarily analytic) functions of two real variables. To this end, (\ref{decomp7}) is first expressed in block matrix notation:
\begin{eqnarray}
&& \left[\begin{array}{c|c}
   \Delta\Delta F_{\alpha\beta} & \Delta\Delta F_{\alpha z}\\
\hline
   \Delta\Delta F_{z\alpha} & \Delta\Delta F
 \end{array} 
 \right]=   \left[\begin{array}{c|c}
   \eta^\star_{\alpha\beta} & \eta^\star_{\alpha z}\\
\hline
   \eta^\star_{z\alpha} & \eta^\star_{zz}
 \end{array} 
 \right]\nonumber\\
&=& \left[\begin{array}{c|c}
\begin{array}{cc}\partial^2_y\mathcal{E}^\star_{zz}&-\partial_x\partial_y\mathcal{E}^\star_{zz}\\
-\partial_x\partial_y\mathcal{E}^\star_{zz} &\partial^2_x\mathcal{E}^\star_{zz}\end{array} & \begin{array}{c}\partial_y\left(\partial_x\mathcal{E}^\star_{yz}-\partial_y\mathcal{E}^\star_{xz}\right)\\
-\partial_x\left(\partial_x\mathcal{E}^\star_{yz}-\partial_y\mathcal{E}^\star_{xz}\right)\end{array}\\
\hline
\begin{array}{cc}
\begin{array}{c}\partial_y\left(\partial_x\mathcal{E}^\star_{yz}\right.\\
\left.-\partial_y\mathcal{E}^\star_{xz}\right)\end{array}
&
\begin{array}{c}-\partial_x\left(\partial_x\mathcal{E}^\star_{yz}\right.\\
\left.-\partial_y\mathcal{E}^\star_{xz}\right)\end{array}
\end{array} &
\begin{array}{c}\partial_x\left(\partial_x\mathcal{E}^\star_{yy}-\partial_y\mathcal{E}^\star_{xy}\right)\\\partial_y\left(\partial_x\mathcal{E}^\star_{xy}-\partial_y\mathcal{E}^\star_{xx}\right)\end{array}
\end{array} 
 \right]\label{decompstrain}
\end{eqnarray} 
and every block equation is separately solved.
\nl
$\blacktriangle$ {First block} ($\eta^\star_{\alpha\beta}$). By the compatibility condition (\ref{eqcomplane}) outside $\lr_0$, it results that $\mathcal{E}^\star_{zz}$ is linear on the open and connected defect-free region $\Om_{z_0}\setminus\lr_0$ and can be prolonged by a linear function on $\Om_{z_0}$. Since $F_{\alpha\beta}=0$ is a solution of $\Delta\Delta F_{\alpha\beta}=0$ on $\Om_{z_0}$, by (\ref{decompstrain}) an admissible $\strains_{zz}$ is
\begin{eqnarray}
 \mathcal{E}^{\star s}_{zz}=0.\label{strain_zz}
\end{eqnarray} 
$\blacktriangle$ {Second block} ($\eta^\star_{\alpha z}$). As $\overline\partial_z\omega^\star_z=\partial_x\mathcal{E}^\star_{yz}-\partial_y\mathcal{E}^\star_{xz}$, it is convenient to seek a solution of
\begin{eqnarray}
 \Delta\Delta\left(F_{xz}+iF_{yz}\right)=-i\left(\partial_x+i\partial_y\right)\overline\partial_z\omega^\star_z\label{FD1}.
\end{eqnarray} 

According to assumption \ref{assfranktens}, $\overline\partial_z\omega^\star_z$ decomposes as follows:
\begin{eqnarray}
\overline\partial_z\omega^\star_z=\sum_{i\in\mathcal{I}}c^{(i)}\delta^{(i)}+K,\label{comp_2}
\end{eqnarray} 
where the absolutely continuous part of the now finite Radon measure $\overline\partial_z\omega^\star_z$ shows to be the constant $K$ appearing in the compatibility condition (\ref{eqcomplane}), while its singular part is purely concentrated. Moreover, the sum of the weights $c^{(i)}$ is absolutely convergent by lemma \ref{L1} when $\Om_{z_0}$ is bounded. 
So (\ref{FD1}) develops as
\begin{eqnarray}
 \Delta\Delta\left(F_{xz}+iF_{yz}\right)=-i\left(\partial_x+i\partial_y\right)\sum_{i\in\mathcal{I}}c^{(i)}\delta^{(i)}\label{FD2}.
\end{eqnarray}

Then, since $\Delta\Delta$ rewrites as $\left(\partial_x+i\partial_y\right)^2\left(\partial_x-i\partial_y\right)^2$, it suffices to solve
\begin{eqnarray}
\left(\partial_x-i\partial_y\right)\left(F_{xz}+iF_{yz}\right)=-i\mathcal{F},\label{comp_1}
\end{eqnarray} 
with $\mathcal{F}$ a solution of
\begin{eqnarray}
\Delta\mathcal{F}=\sum_{i\in\mathcal{I}}c^{(i)}\delta^{(i)}.\label{FD3}
\end{eqnarray}
To solve this system, observe that (\ref{comp_1}) develops as 
\begin{eqnarray}
\left\{
\begin{array}{ccc}
\partial_xF_{yz}-\partial_yF_{xz}&=&-\mathcal{F}\label{comp_1b},\\
\partial_xF_{xz}+\partial_yF_{yz}&=&0,\label{comp_2b}\end{array}\right.
\end{eqnarray} 
where an acceptable $W^{1,1}(\Om_{z_0})$ field $\mathcal{F}$ satisfying (\ref{FD3}) is
\begin{eqnarray}
\mathcal{F}(\chi)=\sum_{i\in\mathcal{I}}\frac{c^{(i)}}{2\pi}\log r^{(i)},\label{solF_1}
\end{eqnarray} 
using the notations $\chi=(x,y)$, $\lr_0=\{\hat\chi^{(i)}=(x^{(i)},y^{(i)}),\ i\in\mathcal{I}\}$ and $r^{(i)}:=|\chi-\hat\chi^{(i)}|$. Hence, by (\ref{decomp4}) an admissible solenoidal strain belonging to $L^1(\Om_{z_0})$ is
\begin{eqnarray}
\mathcal{E}^{\star s}_{xz}&=&\partial_y\left(\partial_xF_{yz}-\partial_yF_{xz}\right)=-\partial_y\mathcal{F},\label{strain_xz}\\
 \mathcal{E}^{\star s}_{yz}&=&-\partial_x\left(\partial_xF_{yz}-\partial_yF_{xz}\right)=\partial_x\mathcal{F}.
\label{strain_yz}
\end{eqnarray} 
$\blacktriangle$ {Third block} ($\eta^\star_{zz}$). Recall first that $\overline\partial_\beta\omega^\star_z=\epsilon_{\alpha\gamma}\partial_\alpha\mathcal{E}^\star_{\gamma\beta}\nonumber$ 
and that
\begin{eqnarray}
\eta^\star_{zz}=\partial_\alpha\left(\partial_\alpha\mathcal{E}^\star_{\beta\beta}-\partial_\beta\mathcal{E}^\star_{\alpha\beta}\right)=\epsilon_{\alpha\beta}\partial_\alpha\overline\partial_\beta\omega^\star_z,\nonumber
\end{eqnarray} 
in such a way that the problem is to solve
\begin{eqnarray}
\Delta\Delta F=\partial_x\overline\partial_y\omega^\star_z-\partial_y\overline\partial_x\omega^\star_z=\Re\{i(\partial_x+i\partial_y)(\overline\partial_x\omega^\star_z-i\overline\partial_y\omega^\star_z)\}\label{FD4}
\end{eqnarray}  
where, according to assumption \ref{assfranktens}, $\overline \partial_\alpha\omega^\star_z$ develops as
\begin{equation}
\overline \partial_\alpha\omega^\star_z=\sum_{i\in\mathcal{I}}c_\alpha^{(i)}\delta^{(i)}+f_\alpha \label{decompfrank}
\end{equation} 
with absolutely convergent sums of the weights $c_\alpha^{(i)}$ by lemma \ref{L1} when $\Om_{z_0}$ is bounded and for some functions $f_\alpha$ whose curl is here a finite Radon measure, which must be concentrated on $\lr_0$ to ensure the compatibility of $\strain_{\alpha\beta}$ outside the defect set. So, (\ref{FD4}) rewrites as  
\begin{eqnarray}
 \Delta\Delta F = \Re\left\{i(\partial_x+i\partial_y)\left(\sum_{i\in\mathcal{I}}\left(c_x^{(i)}-ic_y^{(i)}\right)\delta^{(i)}\right)\right\}+\partial_x f_y-\partial_y f_x.\nonumber
\end{eqnarray}

Now, in view of the properties of $f_\alpha$ resulting from assumption \ref{assfranktens}, the last terms write as
\begin{eqnarray}
\partial_x f_y-\partial_y f_x =  \sum_{i\in\mathcal{I}}C^{(i)}\delta^{(i)}\label{FD5},
\end{eqnarray}
where the sum of the weights is absolutely convergent. Eventually, using the distributional identity
\begin{eqnarray}
 \partial_x\left(\frac{\delta x^{(i)}}{r^{(i)2}}\right)-\partial_y\left(\frac{\delta y^{(i)}}{r^{(i)2}}\right)=2\pi\delta ^{(i)},\nonumber
\end{eqnarray}
with the notation $\delta x^{(i)}=x-\hat x^{(i)}, \delta y^{(i)}=y-\hat y^{(i)}$, (\ref{FD4}) can be written in the form
\begin{eqnarray}
 \Delta\Delta F = \Re\left\{i(\partial_x+i\partial_y)\left(\sum_{i\in\mathcal{I}}\left(c_x^{(i)}-ic_y^{(i)}\right)\delta^{(i)}
-\sum_{i\in\mathcal{I}}\frac{C^{(i)}}{2\pi r^{(i)2}}\left(y^{(i)}+ix^{(i)}\right)\right)\right\}.\nonumber\\\quad\label{FD6}
\end{eqnarray}

A particular solution of (\ref{FD6}) is provided by solving
\begin{eqnarray}
 (\partial_x+i\partial_y)(\partial_x-i\partial_y)^2(F+iH)=i\left(\sum_{i\in\mathcal{I}}\left(c_x^{(i)}-ic_y^{(i)}\right)\delta^{(i)}
\right.&&\nonumber\\
\left.-\sum_{i\in\mathcal{I}}\frac{C^{(i)}}{2\pi r^{(i)2}}\left(y^{(i)}+ix^{(i)}\right)\right),&&\label{primal}
\end{eqnarray} 
with $H$ an additional unknown. This latter equation is equivalent to the system
\begin{eqnarray}
(\partial_x-i\partial_y)(F+iH)=\mathcal{G}\quad\mbox{on}\quad \Om_{z_0},&&\label{zz_1}\\
 \Delta\mathcal{G}=i(\overline\partial_x\omega^\star_z-i\overline\partial_y\omega^\star_z)\quad\mbox{on}\quad \Om_{z_0},&&\label{zz_1a}
\end{eqnarray} 
which can be easily solved. In a first step, a particular solution of (\ref{zz_1a}) is given by
\begin{eqnarray}
\mathcal{G}&=&\mathcal{G}_1+\mathcal{G}_2,\nonumber\\
\mathcal{G}_1&=&\sum_{i\in\mathcal{I}}\left(c_y^{(i)}+ic_x^{(i)}\right)\frac{\log r^{(i)}}{2\pi},
\label{un}\\
\mathcal{G}_2&=&\sum_{i\in\mathcal{I}}C^{(i)}\left(\delta x^{(i)}-i\delta y^{(i)}\right)\frac{\log r^{(i)}}{4\pi},\label{deux}
\end{eqnarray} 
with both $\mathcal{G}_1$ and $\mathcal{G}_2$ belonging to $W^{1,1}(\Om_{z_0})$. In a second step, 
 (\ref{zz_1}) is simply rewritten as
\begin{eqnarray}
\partial_xF+\partial_yH=\Re\{\mathcal{G}\}\quad\mbox{and}\quad
 \partial_xH-\partial_yF=\Im\{\mathcal{G}\}  ,   \label{zz_1b}
\end{eqnarray} 
whose solution $F=F_1+F_2$ and $H=H_1+H_2$ is given by
\begin{eqnarray}
F_1=\partial_x\psi_1+\partial_y\varphi_1\quad&\mbox{and}&\quad H_1=\partial_y\psi_1-\partial_x\varphi_1,\label{ac1}\\
F_2=\partial_x\psi_2+\partial_y\varphi_2 \quad&\mbox{and}&\quad H_2=\partial_y\psi_2-\partial_x\varphi_2,\label{c1}
\end{eqnarray} 
for some gauge fields $\psi_1,\varphi_1,\psi_2,\varphi_2$ satisfying the equations
\begin{eqnarray}
 \Delta\psi_1=\Re\{\mathcal{G}_1\}\quad&\mbox{and}&\quad\Delta\varphi_1=-\Im\{\mathcal{G}_1\},\label{ac2}\\
\Delta\psi_2=\Re\{\mathcal{G}_2\}\quad&\mbox{and}&\quad \Delta\varphi_2=-\Im\{\mathcal{G}_2\}.\label{c2}
\end{eqnarray} 

Particular solutions  of (\ref{ac2}) belonging to $W^{3,1}(\Om_{z_0})$ are 
\begin{eqnarray}
 \psi_1=\sum_{i\in\mathcal{I}}c_y^{(i)}r^{(i)2}\frac{\log r^{(i)}-1}{8\pi}\quad\mbox{and}\quad
 \varphi_1=-\sum_{i\in\mathcal{I}}c_x^{(i)}r^{(i)2}\frac{\log r^{(i)}-1}{8\pi},\nonumber
\end{eqnarray} 
in such a way that
\begin{eqnarray}
F_1(x,y)=\sum_{i\in\mathcal{I}}\frac{(2\log r^{(i)}-1)}{8\pi}\left(c^{(i)}_y\delta x^{(i)}-c^{(i)}_x\delta	 y^{(i)}\right)&&\label{Fzz}
\end{eqnarray} 
belongs to $W^{2,1}(\Om_{z_0})$, thereby defining the solenoidal strain $\mathcal{E}^{\star s,1}_{\alpha\beta}$:
\begin{eqnarray}
[\mathcal{E}^{\star s,1}_{\alpha\beta}]:=\left[\begin{array}{cc}\partial^2_yF_1&-\partial_x\partial_yF_1\\
-\partial_x\partial_yF_1 &\partial^2_xF_1\end{array}\right],\label{strain_alphabeta_1}
\end{eqnarray} 
which belongs to $L^1(\Om_{z_0})\cap\mathcal{C}^\infty(\Om_{z_0}\setminus{\lr_0})$. Similarly, particular solutions of (\ref{c2}) belonging to $W^{3,1}(\Om_{z_0})$ are given by 
\begin{eqnarray}
 \psi_2&=&\sum_{i\in\mathcal{I}}C^{(i)}\delta x^{(i)}r^{(i)2}\frac{\log r^{(i)}-3/4}{32\pi},\label{FD7}\\
 \varphi_2&=&\sum_{i\in\mathcal{I}}C^{(i)}\delta y^{(i)}r^{(i)2}\frac{\log r^{(i)}-3/4}{32\pi},\label{FD8}
\end{eqnarray} 
and hence
\begin{eqnarray}
F_2(x,y)=\sum_{i\in\mathcal{I}}\frac{C^{(i)}}{16\pi}r^{(i)2}\left(2\log r^{(i)}-1)\right)&&\label{Fzz2}
\end{eqnarray} 
also belongs to $W^{2,1}(\Om_{z_0})$, defining the solenoidal strain $\mathcal{E}^{\star s,2}_{\alpha\beta}$:
\begin{eqnarray}
[\mathcal{E}^{\star s,2}_{\alpha\beta}]:=\left[\begin{array}{cc}\partial^2_yF_2&-\partial_x\partial_yF_2\\
-\partial_x\partial_yF_2 &\partial^2_xF_2\end{array}\right],\label{strain_alphabeta_2}
\end{eqnarray} 
which belongs to $L^1(\Om_{z_0})\cap\mathcal{C}^\infty(\Om_{z_0}\setminus{\lr_0})$.
\nl\nl
$\blacktriangle$ {Summary.} The solenoidal part of the strain is the tensor $\mathcal{E}^{\star s}_{mn}$,
\begin{eqnarray}
 \mathcal{E}^{\star s}_{mn}=\mathcal{E}^{\star s}_{zz}\delta_{mz}\delta_{nz}+\mathcal{E}^{\star s}_{\alpha z}\left(\delta_{mz}\delta_{n\alpha}+\delta_{m\alpha}\delta_{nz}\right)+\left(\mathcal{E}^{\star s,1}_{\alpha\beta}+\mathcal{E}^{\star s,2}_{\alpha\beta}\right)\delta_{m\alpha}\delta_{n\beta},\label{strainsolenoidal}
\end{eqnarray} 
where $\mathcal{E}^{\star s}_{zz}, \mathcal{E}^{\star s}_{\alpha z}, \mathcal{E}^{\star s,1}_{\alpha\beta}$ and $\mathcal{E}^{\star s,2}_{\alpha\beta}$ are given by (\ref{strain_zz}), (\ref{solF_1}), (\ref{strain_xz}), (\ref{strain_alphabeta_1}) and (\ref{strain_alphabeta_2}), respectively, and all belong to $L^1(\Om_{z_0})$, observing that the weights $c^{(i)}, c_\alpha^{(i)}$ and $C^{(i)}$ defining the intermediate expressions $\mathcal{F}, F_1$ and $F_2$ in (\ref{FD3}), (\ref{Fzz}) and (\ref{Fzz2}) are associated with absolutely convergent series:
\begin{eqnarray}
 \sum_{i\in\mathcal{I}}|c^{(i)}|<\infty,\quad\sum_{i\in\mathcal{I}}||c^{(i)}_\alpha||<\infty\quad \text{and}\quad \sum_{i\in\mathcal{I}}|C^{(i)}|<\infty.\label{sumc}
\end{eqnarray}
Therefore  $\mathcal{E}^{\star s}_{mn}$ belongs to $L^1(\Om_{z_0})$, is solenoidal and satisfies (\ref{strain_decomp}). Extension of the proof to unbounded domains $\Om_{z_0}$ is immediate.
{\hfill $\square$}

The coefficients $c^{(i)},c^{(i)}_\alpha$ and $C^{(i)}$ will show to be the Burgers and Frank vectors of screw and edge dislocations and wedge disclinations, respectively.
\begin{Rem}
The hypothesis provided by assumption \ref{assfranktens} that $\overline\partial_m\omega^\star_z$ has an absolutely continuous part whose curl is a Radon measure is a request to make the proof in the case of a countable set of line defects. Indeed, when the $2D$ defect set $\lr_0$ has accumulation points in $\Om_{z_0}$, a complex distributional behaviour can take place near these points which forbids geting the proof if a sufficiently strong hypothesis is not introduced to account for a possibly countable number of disclinations on the sole basis of the strain field properties. More tractable hypotheses on $\overline\partial_\alpha\omega^\star_z$ itself (and not its curl) are currently under investigation. 
\end{Rem}

As a $1D$ example to illustrate the above difficulty, the function 

\begin{equation}
F=\displaystyle\sum_{i\in\mathcal{I}=\NN_0}C^{(i)}(H^0-H^{(i)})\nonumber
\end{equation} 
with $H^{(i)}=H(x-\hat x^{(i)}),\ \hat x^{(i)}=1/i,\ H^0=H(x)$
and $H$ the step function, may correspond to an $L^1_{loc}$ converging series even if the sum of the weights $C^{(i)}$ diverges. To show this, it suffices to select appropriate $C^{(i)}$ such that the partial sums defining $F$ are all enclosed between the $L^1_{loc}$ functions $G(x)$ and $-G(x)$, with $G(x)=\log \left(\left(1+x\right)/x\right)$ for $x >0$ and $G(x)=0$ for $x\leq 0$. The Lebesgue dominated convergence theorem then shows that $F\in L^1_{loc}$, in such a way that the distributional derivative of $F$, which cannot be the diverging series $-\displaystyle\sum_{i\in\mathcal{I}}C^{(i)}\delta^{(i)}$, exhibits a special behaviour near the origin to recover convergence. Similar effects take place in $2D$ and appropriate assumptions are then necessary to obtain (\ref{FD5}).

\subsection{Second representation theorem of a $2D$ incompatibile strain}
\label{sec:straindecomp2}

This section provides a further decomposition of the strain, since the solenoidal part is itself decomposed into an everywhere compatible part and another smooth part outside from the defect set $\Lr$.

\begin{theorem}[analysis of the singular part of the strain decomposition]\label{straindecomp2}
 Let the strain and the Frank tensor satisfy assumptions \ref{asstrain} and \ref{assfranktens}, and the dislocation set be defined according to definition \ref{lines2D}. Then the solenoidal component of the strain satisfies the following decomposition:
\begin{eqnarray}
 \mathcal{E}^{\star s}_{mn}= \mathcal{E}'^{\star c}_{mn}+ E^\star_{mn},\label{strain_decomp_2}
\end{eqnarray} 
where $\mathcal{E}'^{\star c}_{mn}\in L^1_{loc}(\Om_{z_0})$ is compatible on $\Om$ and where
\begin{equation}
E^\star_{mn}=\sum_{i\in\mathcal{I}}E^{\star (i)}_{mn}\in L^1_{loc}(\Om),\label{strain_somme}
\end{equation} 
with  $E^{\star (i)}_{mn} (i\in\mathcal{I})$ smooth and compatible on $\Om\setminus L^{(i)}$. Moreover, the Frank tensor part $\epsilon_{kpn}\partial_p E^\star_{mn}$ is smooth on $\OmLr$.
\end{theorem}
{\bf Proof.} As in the previous proof, since the strain is independent of $z$, it suffices to work on the $2D$ domain $\Om_{z_0}$ which is again assumed to be bounded without loss of generality.
\nl\nl
$\blacktriangle$ {$\strains_{\alpha z}$ components.} Following decomposition (\ref{strain_decomp}), write
\begin{eqnarray}
\mathcal{E}^{\star s}_{xz}&=&\partial_y\left(\partial_xF_{yz}-\partial_yF_{xz}\right)=E^\star_{xz}+\mathcal{E'}^{\star c}_{xz},\nonumber\\
 \mathcal{E}^{\star s}_{yz}&=&-\partial_x\left(\partial_xF_{yz}-\partial_yF_{xz}\right)=E^\star_{yz}+\mathcal{E'}^{\star c}_{yz},
\nonumber
\end{eqnarray} 
with vanishing $\mathcal{E}'^{\star c}_{xz}$ and $\mathcal{E}'^{\star c}_{yz}$:
\begin{equation}
 \mathcal{E}'^{\star c}_{xz}=\mathcal{E}'^{\star c}_{yz}=0.\nonumber
\end{equation}
Hence, according to (\ref{strain_xz}) and (\ref{strain_yz}), $E^\star_{xz}$ and $E^\star_{yz}$ are sums of screw dislocations:
\begin{eqnarray}
E^\star_{xz}=-\sum_{i\in\mathcal{I}}\frac{c^{(i)}}{2\pi r^{(i)2}}\delta y^{(i)}\quad\mbox{and}\quad  E^\star_{yz}=\sum_{i\in\mathcal{I}}\frac{c^{(i)}}{2\pi r^{(i)2}}\delta x^{(i)},\label{strain_xyza}
\end{eqnarray} 
which from (\ref{sumc}) belong to $L^1(\Om_{z_0})\cap\mathcal{C}^\infty(\Om_{z_0}\setminus{\lr_0})$. The relations (\ref{strain_xyza}) also show the smoothness of the Frank tensor parts $\epsilon_{\alpha\beta}\partial_\alpha E^\star_{z\beta}$ and $\epsilon_{\beta\gamma}\partial_\gamma E^\star_{\alpha z}$ outside $\lr_0$.
\nl\nl
$\blacktriangle$ {$\strains_{zz}$ and $\mathcal{E}^{\star s,1}_{\alpha\beta}$ components.} By (\ref{strain_zz}) the expression 
\begin{eqnarray}
E^\star_{zz}:=\mathcal{E}^{\star s}_{zz}=0\label{elections}
\end{eqnarray}  
exhibits the form (\ref{strain_somme}) and provides Frank tensor parts $\epsilon_{\alpha\gamma}\partial_\gamma E^\star_{zz}$ which identically vanish on $\Om_{z_0}$. On the other hand, it can be checked that on $\Om_{z_0}\setminus{\lr_0}$:
\begin{eqnarray}
[\mathcal{E}^{\star s,1}_{\alpha\beta}]=\sum_{i\in\mathcal{I}}\frac{c_y^{(i)}}{4\pi r^{(i)2}}\left[\begin{array}{cc}\delta x^{(i)}(1-2\frac{\delta y^{(i)2}}{r^{(i)2}}) & -\delta y^{(i)}(1-2\frac{\delta x^{(i)2}}{r^{(i)2}})\\
-\delta y^{(i)}(1-2\frac{\delta x^{(i)2}}{r^{(i)2}})& \delta x^{(i)}(1+2\frac{\delta y^{(i)2}}{r^{(i)2}})\end{array}\right]\nonumber\\
-\sum_{i\in\mathcal{I}}\frac{c_x^{(i)}}{4\pi r^{(i)2}}\left[\begin{array}{cc}\delta y^{(i)}(1+2\frac{\delta x^{(i)2}}{r^{(i)2}}) & -\delta x^{(i)}(1-2\frac{\delta y^{(i)2}}{r^{(i)2}})\\
-\delta x^{(i)}(1-2\frac{\delta y^{(i)2}}{r^{(i)2}}) & \delta y^{(i)}(1-2\frac{\delta x^{(i)2}}{r^{(i)2}})\end{array}\right].\label{E1}
\end{eqnarray} 
Define then in the decomposition (\ref{strain_decomp_2}), (\ref{strain_somme}) of $\mathcal{E}^{\star s,1}_{\alpha\beta}$ the components $E^{\star 1}_{\alpha\beta}$ and $\mathcal{E'}^{\star c,1}_{\alpha\beta}$ as follows:
\begin{eqnarray}
[E^{\star 1}_{\alpha\beta}]:=-\sum_{i\in\mathcal{I}}\frac{c^{(i)}_y}{2\pi r^{(i)2}}\left [\begin{array}{cc}\delta x^{(i)}&\delta y^{(i)}\\ \delta y^{(i)}&-\delta x^{(i)}
\end{array} \right ]+\sum_{i\in\mathcal{I}}\frac{c^{(i)}_x}{2\pi r^{(i)2}}\left [\begin{array}{ccc} -\delta y^{(i)}&\delta x^{(i)}\\ \delta x^{(i)}&\delta y^{(i)}
\end{array} \right ],&&\label{E1b}\\
 {[}\mathcal{E'}^{\star c,1}_{\alpha\beta}]:=\sum_{i\in\mathcal{I}}\frac{c_y^{(i)}}{4\pi r^{(i)4}}\left[\begin{array}{cc}\delta x^{(i)}(\delta y^{(i)2}+3\delta x^{(i)2})&\delta y^{(i)}(\delta y^{(i)2}+3\delta x^{(i)2})\\
\delta y^{(i)}(\delta y^{(i)2}+3\delta x^{(i)2}) &\delta x^{(i)}(-\delta x^{(i)2}+\delta y^{(i)2})\end{array}\right]&&\nonumber\\
-\sum_{i\in\mathcal{I}}\frac{c_x^{(i)}}{4\pi r^{(i)4}}\left[\begin{array}{cc}\delta y^{(i)}(\delta x^{(i)2}-\delta y^{(i)2})&\delta x^{(i)}(\delta x^{(i)2}+3\delta y^{(i)2})\\
\delta x^{(i)}(\delta x^{(i)2}+3\delta y^{(i)2}) &\delta y^{(i)}(\delta x^{(i)2}+3\delta y^{(i)2})\end{array}\right].&&\label{E3}
\end{eqnarray} 
Calculations show that $\mathcal{E'}^{\star c,1}_{\alpha\beta}$ is the difference between (\ref{E1}) and (\ref{E1b}), is compatible, and also belongs to $L^1(\Om_{z_0})$ by (\ref{sumc}). Moreover $E^{\star 1}_{\alpha\beta}$ is a sum of edge dislocations, whose curl is vanishing and which are smooth on $\Om_{z_0}\setminus\{(\hat x^{(i)},\hat y^{(i)})\}$.
\nl\nl
$\blacktriangle$ {$\mathcal{E}^{\star s,2}_{\alpha\beta}$ components.} Define
\begin{eqnarray}
E^{\star 2}_{\alpha\beta}:=\mathcal{E}^{\star s,2}_{\alpha\beta}\label{election2}
\end{eqnarray}  
with vanishing $\mathcal{E'}^{\star c,2}_{\alpha\beta}$. Calculations show that
\begin{eqnarray}
[E^{\star 2}_{\alpha\beta}]=\sum_{i\in\mathcal{I}}\frac{C^{(i)}}{4\pi}\left[\begin{array}{cc}\log r^{(i)}+\frac{\delta y^{(i)2}}{r^{(i)2}} & -\frac{\delta x^{(i)}\delta y^{(i)}}{r^{(i)2}}\\
-\frac{\delta x^{(i)}\delta y^{(i)}}{r^{(i)2}} & \log r^{(i)}+\frac{\delta x^{(i)2}}{r^{(i)2}}\end{array}\right],\label{E2}
\end{eqnarray} 
in such a way that $E^{\star 2}_{\alpha\beta}$ is a sum of wedge disclinations, which are smooth and of vanishing curl on $\Om_{z_0}\setminus\{(\hat x^{(i)},\hat y^{(i)})\}$ (cf. Part A).
\nl\nl
$\blacktriangle$ In summary, the following solenoidal strain decomposition has been proved:
 \begin{eqnarray}
 \mathcal{E}^{\star s}_{mn}= \mathcal{E}'^{\star c}_{mn}+ E^\star_{mn},\label{strain_decomp_3}
\end{eqnarray} 
where
\begin{eqnarray}
\mathcal{E}'^{\star c}_{mn}:=\mathcal{E}'^{\star c}_{zz}
\delta_{mz}\delta_{nz}+\mathcal{E}'^{\star c}_{\alpha z}(\delta_{m\alpha}\delta_{nz}+\delta_{mz}\delta_{n\alpha })+\left(\mathcal{E}'^{\star c,1}_{\alpha\beta}+\mathcal{E}'^{\star c,2}_{\alpha\beta}\right)\delta_{m\alpha}\delta_{n\beta}
\end{eqnarray} 
is compatible on $\Om$. Moreover, $E^\star_{mn}$ writes as
\begin{eqnarray}
E^\star_{mn}=E^\star_{zz}
\delta_{mz}\delta_{nz}+E^\star_{\alpha z}(\delta_{m\alpha}\delta_{nz}+\delta_{mz}\delta_{n\alpha})+\left(E^{\star 1}_{\alpha\beta}+E^{\star 2}_{\alpha\beta}\right)\delta_{m\alpha}\delta_{n\beta},\label{strain_somme_a}
\end{eqnarray} 
with $E^\star_{zz},E^\star_{\beta z}, E^{\star 1}_{\alpha\beta}$ and $E^{\star 2}_{\alpha\beta}$ defined by (\ref{elections}), (\ref{strain_xyza}), (\ref{E1b}) and (\ref{E2}), and hence $E^\star_{mn}$ together with the related Frank tensor
$\epsilon_{kpn}\partial_p E^\star_{mn}$ is smooth on $\OmLr$ thereby terminating the proof. Extension to unbounded sets $\Om_{z_0}$ is straightforward.
{\hfill $\square$}

\subsection{Applications of the strain decomposition}
\label{sec:appstraindecomp}
In this \S\ref{distrset}\ref{sec:appstraindecomp}, $\mathcal{I}'$ refers to any bounded subset $\{\hat x^{(i)},\ i\in\mathcal{I}'\}$ of $\lr_0$.\nl\nl
$\bullet\ $ \textit{Set of parallel screw disclocations.} Part A and (\ref{strain_xyza}) directly provide the equality $c^{(i)}=B^{\star (i)}_z$ and a vanishing strain compatible part. Moreover according to (\ref{sumc}) the following condition holds:
\begin{eqnarray}
\sum_{i\in\mathcal{I}'}|B^{\star (i)}_z|<\infty.\nonumber
\end{eqnarray} 
$\bullet$ \textit{Set of parallel edge disclocations.} 
From Part A and (\ref{E1})-(\ref{E3}), it turns out that $c^{(i)}_{yz}=B^{\star (i)}_y$ and $c^{(i)}_{xz}=B^{\star (i)}_x$, with according to (\ref{sumc}) the following bounds:
\begin{eqnarray}
 \sum_{i\in\mathcal{I}'}|B^{\star (i)}_x|<\infty\quad\text{and}\quad \sum_{i\in\mathcal{I}'}|B^{\star (i)}_y|<\infty.\nonumber
\end{eqnarray} 
$\bullet$ \textit{Set of parallel wedge disclinations.} The  expression of $[\strain_{ij}]$ given in Part A is
\begin{eqnarray}
\frac{\Om^\star_{z}(1-\nu^{\star})}{4\pi}\left [ \begin{array}{ccc} 1+\log r& 0& 0 \\0&1+\log r & 0  \\ 0 & 0 &0
\end{array} \right ]-\frac{\Om^\star_z(1+\nu^*)}{8\pi}\left [ \begin{array}{ccc} \cos2\theta& \sin2\theta & 0 \\\sin2\theta &-\cos2\theta & 0  \\ 0 & 0 &0
\end{array} \right]\nonumber,
\end{eqnarray} 
which shows to be the sum of 
\begin{eqnarray}
\frac{\Om^\star_z}{4\pi}\left [\begin{array}{ccc}\log r+\sin^2\theta & -\sin\theta\cos\theta&0 \\ -\sin\theta\cos\theta &\log r+\cos^2\theta & 0  \\ 0 & 0 &0
\end{array} \right ]\nonumber,
\end{eqnarray} 
and a compatible part (since the associated Frank tensor part vanishes). Therefore, according to (\ref{E2}), $C^{(i)}=\Om^{\star (i)}_z$ with by (\ref{sumc}) the bounds 
\begin{equation}
\displaystyle \sum_{i\in\mathcal{I}'}|\Om_z^{(i)}|<\infty. \nonumber
\end{equation} 

\subsection{Mesoscopic defect densities in $2D$ incompatible elasticity}
\label{defectdens}

The following theorem expresses the $2D$ mesoscopic incompatibility in terms of the defect invariants for a countable set of dislocations.
\begin{theorem}[main result] \label{mainresultglob}
For a countable set of parallel defect lines $\Lr$ and under assumptions \ref{asstrain} and \ref{assfranktens} and definition \ref{lines2D}, incompatibility as defined by equations (\ref{eta}) and (\ref{eqcomplane}) is the vectorial first order distribution
\begin{eqnarray}
\eta_k^{\star}=\delta_{kz}\eta_z^{\star}+\delta_{k\kappa}\eta_\kappa^{\star},\label{eta_k}\label{hugo}
\end{eqnarray}
where its out-of-plane component is
\begin{eqnarray}
\eta^\star_z=\sum_{i\in\mathcal{I}}\left(\Om^{\star (i)}_z\delta_{L^{(i)}}+\epsilon_{\alpha\gamma}\left(B_\gamma^{\star (i)}+\epsilon_{\beta\gamma}(\hat x^{(i)}_\beta-x_{0\beta})\Om^{\star (i)}_z\right)\partial_\alpha\delta_{L^{(i)}}\right)\label{final1}
\end{eqnarray}
while its in-plane components are
\begin{eqnarray}
\eta^\star_\kappa=\frac{1}{2}\epsilon_{\kappa\alpha}\sum_{i\in\mathcal{I}}B^{\star (i)}_z\partial_\alpha\delta_{L^{(i)}},\label{final2}
\end{eqnarray}
and where $x_0\in\Om$ is a selected reference point.
\end{theorem}
{\bf Proof.} 
With use of theorems \ref{straindecomp} and \ref{straindecomp2}, the $L^1(\Om)$ strain decomposes as:
\begin{eqnarray}
 \mathcal{E}^\star_{mn}= e^\star_{mn}+E^\star_{mn},\label{strain_decomp_2b}
\end{eqnarray} 
where $e^\star_{mn}$ is compatible on $\Om$ and where 
\begin{eqnarray}
 E^\star_{mn}=\sum_{i\in\mathcal{I}}E^{\star(i)}_{mn}
\end{eqnarray} 
is smooth away from the defect set $\displaystyle\Lr=\bigcup_{i\in\mathcal{I}}L^{(i)}$.

Now, the local strain assumption of Part A is satisfied by each individual $E^{\star(i)}_{mn}$. Then, the Frank and Burgers vectors of each isolated defect line $L^{(i)}$ are defined according to theorem \ref{displ} and definition \ref{Burgers}. In a next step, the strain contributions $E^{\star (i)}_{mn}$ associated with these isolated defect lines are removed from the decomposition of the strain $\strain_{mn}$ provided by (\ref{strain_decomp}), (\ref{strain_decomp_2}) and (\ref{strain_somme}), as allowed by the absolute convergence (\ref{sumc}) of all their weight series. This operation defines a strain reminder whose defect lines are the accumulation lines (or the so-called derived set) of the initial defective set $\Lr$, and the extraction procedure of isolated defect lines is repeated on this derived set, and then repeated again by transfinite recursion until the Frank and Burgers vectors are defined for each line of $\Lr$ (and not only for the isolated lines). Finally, applying the main theorem of Part A to each $L^{(i)}$, summing the results on $\mathcal{I}\subset\NN$ and recalling that both $\mathcal{E}^{\star c}_{mn}$ and $\mathcal{E}'^{\star c}_{mn}$  provide vanishing contributions to incompatibility, the proof is achieved.
{\hfill $\square$}
\nl

The remaining of this section consists in a reformulation of the above result in terms of the defect density tensors
 $\Theta^\star_{ik},\Lambda^\star_{ik},\alpha^\star_{ik}$ and $\kappa^\star_{ik}$ which, in $2D$, simplify in the  $\Theta^\star_k,\Lambda^\star_k,\alpha^\star_k$ and $\kappa^\star_{ij}$ tensors defined by (\ref{disclindens1})-(\ref{dislocdens2b}). However, the sums are now performed on a countable ensemble $\mathcal{L}$ of rectilinear dislocations, and $\Theta^\star_k$ and $\Lambda^\star_k$ are Radon measures in view of the inequalities
\begin{eqnarray}
 \sum_{i\in\mathcal{I}'}|\Om^{\star (i)}_z|<\infty,\quad \sum_{i\in\mathcal{I}'}\parallel B^{\star (i)}_k\parallel<\infty,\label{bounded}
\end{eqnarray} 
where $\mathcal{I}'$ refers to any bounded subset $\{\hat x^{(i)},\ i\in\mathcal{I}'\}$ of $\lr_0$
\begin{theorem}\label{kronerident}
For a countable set $\Lr$ of parallel defect lines and under assumptions \ref{asstrain} and \ref{assfranktens} and definition \ref{lines2D}, the mesoscopic strain incompatibility writes as
\begin{eqnarray}
\eta^\star_k=\delta_{zk}\Theta^\star_z+\epsilon_{\alpha\beta}\partial_\alpha\kappa_{k\beta}^\star,\label{kroner1}
\end{eqnarray}
or equivalently as 
\begin{eqnarray}
\eta^\star_k=\delta_{zk}\Theta^\star_z+\epsilon_{k\alpha l}\partial_\alpha\kappa^\star_{zl}.\label{FD11}
\end{eqnarray}
\end{theorem}
{\bf Proof.} 
Consider any straight dislocation $L^{(i)}\in\mathcal{L}$ passing by $\hat x^{(i)}=(\hat x^{(i)}_\beta,z_0)$. From theorem \ref{mainresultglob}, the associated incompatibility writes as
\begin{eqnarray}
\eta^{\star (i)}_k=\delta_{kz}\left(\Om^{\star (i)}_z\delta_{L^{(i)}}+\epsilon_{\alpha\gamma}\left(B^{\star (i)}_\gamma+\epsilon_{\beta\gamma}(\hat x^{(i)}_\beta-x_{0\beta})\Om^{\star (i)}_z\right)\partial_\alpha\delta_{L^{(i)}}\right)&&\nonumber\\
+\delta_{k\kappa}\frac{1}{2}\epsilon_{\kappa\alpha}B^{\star (i)}_z\partial_\alpha\delta_{L^{(i)}}.&&\label{translat}
\end{eqnarray}
Taking into account (\ref{disclindens1}), (\ref{dislocdens2}), (\ref{KR3}) and (\ref{KR2}) for a single line, and the relation
\begin{eqnarray}
\partial_\alpha\left((x_\beta-x_{0\beta})\delta_{L^{(i)}}\right)=\partial_\alpha\left((\hat x_\beta^{(i)}-x_{0\beta})\delta_{L^{(i)}}\right)=(\hat x_\beta^{(i)}-x_{0\beta})\partial_\alpha\delta_{L^{(i)}},\nonumber
\end{eqnarray}
it results that $\eta^{\star (i)}_k$ can be written in the formulations (\ref{kroner1}) or (\ref{FD11}), and the result follows after summation over $\mathcal{I}$.
{\hfill $\square$}

\section{Displacement and rotation fields in $2D$ incompatible elasticity at mesoscopic scale}\label{multi_inc}
\subsection{Position of the problem}\label{pospbl}
The principal objective of the present work is to pave the way for a mathematically rigourous treatment of dislocation homogenization from meso- to macro-scale (Van Goethem and Dupret 2009$b$). To this end, this section elucidates the link between incompatibility (expressed as a combination of the concentrated defect densities) and the multivalued displacement and rotation fields. For the sake of completeness, most results are expressed using complete $3D$ tensor components (with latin indices, cf. Part A) under the hypothesis of a $2D$ strain field (whose components do not depend on $z$).

On the one hand, the mesoscopic fields $\Theta^\star_k=\Theta^\star_{zk}, \Lambda^\star_k=\Lambda^\star_{zk}, \eta^\star_k=\eta^\star_{zk}=\eta^\star_{kz}$ and the contortion $\kappa^\star_{ij}$ are concentrated distributions on the defect lines which, as shown in Part A and \S \ref{distrset}, provide all the information on the dislocation and disclination densities and the strain incompatibility. 

On the other hand, by theorem \ref{displ} the rotation field is a multivalued function of index $1$ obtained on $\Om_\Lr$ by line integration of $\partial_l^{(s)} \omega^\star_k=\overline\partial_l\omega^\star_k=\epsilon_{kpq}\partial_p\strain_{ql}$ (cf. (\ref{delta_m_a})) while the displacement field $u^\star_k$ is a multivalued function of index-$2$ obtained on $\Om_\Lr$ by recursive line integration of $\partial_j^{(s)}\partial_l^{(s)} u^\star_k=\partial_j^{(s)}\left(\strain_{kl}+\omega_{kl}^\star\right)$. The Frank tensor was introduced as a distributional field $\overline\partial_l\omega^\star_k$ defined on the entire $\Om$, which coincides with $\partial_l^{(s)} \omega^\star_k$ on every defect-free region and to which additional distributional terms are added to let this tensor be related to the strain gradient by (\ref{delta_m_a}). 

Direct analysis shows that displacement is not the most appropriate vector field to describe the dislocations since the Burgers field $b^\star_k$ defined from (\ref{delta_lb_i}) by line integration on $\Om_\Lr$ of $\partial_l^{(s)} b^\star_k=\overline\partial_lb^\star_k$ exhibits more useful properties. In particular, $b^\star_k$ is a multivalued field of index $1$ (compared to the less tractable index-$2$ multivaluedness of displacement) and the properties of the Frank tensor and vector $\overline\partial_l\omega^\star_k$ and $\omega^\star_k$, and of the Burgers tensor and vector $\overline\partial_l b^\star_k$ and $b^\star_k$, show a clear analogy.

Both the Burgers and the rotation field are defined by means of a Riemann foliation $F$ (cf. Part A) in the sense that mappings of the following kind exist:
\begin{eqnarray}
 \OmLr\stackrel{\mathcal{P}}{\longleftarrow} F \stackrel{\omega^\star_k,b_k^\star}{\longrightarrow} \RR^3\nonumber
\end{eqnarray} 
where $F:=\{(x,\#C)$ for every $x\in\OmLr$ and every curve $C$ joining $x_0$ to $x$, with $\#C$ the equivalence class of all curves homotopic to $C$ in $\OmLr\}$
while $\mathcal{P}$ is the projection of $F$ onto $\OmLr$, in such a way that $\mathcal{P}(x,\#C)=x$. Accordingly, the relationships between the multivalued fields $\omega^\star_k$ and $b^\star_k$ (defined on $F$ together with the projection $\mathcal{P}$), and the distributional fields $\overline\partial_l\omega^\star_k$ and $\overline\partial_l b^\star_k$ (defined on $\Om$) are very similar.

Careful analysis however reveals an apparent contradiction between the expected meanings of the Frank and Burgers tensors and their mathematical properties. Theorem \ref{mainresultglob} first shows that in the absence of disclinations ($\Om^{\star (i)}_z=0$) but in the presence of dislocations, the incompatibility $\eta^\star_k$ does not vanish (this resulting from non-vanishing coefficients multiplying the Dirac mass derivatives ($\partial_\alpha\delta_{L^{(i)}}$) in (\ref{final1}) and (\ref{final2})). Since by (\ref{sup1}) incompatibility is the curl of the Frank tensor, the latter is not curl-free and surprinsingly cannot be the distributional gradient of a single-valued rotation field $\omega^\star_k$ in the absence of rotational defects (a situation where $\omega^\star_k$ is expected to exist in the entire domain $\Om$ and not only on $\Om_\Lr$).

Secondly, the link between the Burgers tensor and vector has also to be clarified, but the situation is slightly different since no pure disclinations exist while there are pure dislocations. Indeed, it should be noticed that the Burgers vectors of the defect lines depend on the reference point $x_0$ in case their Frank tensor does not vanish. More precisely, when $x_0$ is changed to $x'_0$, the following transformation rule applies on each defect line $L^{(i)}$ in the general $3D$ case:
\begin{equation}
 B^{\star '(i)}_k= B^{\star (i)}_k+\epsilon_{klm}\Om^{\star (i)}_l(x'_{0m}-x_{0m}),\nonumber
\end{equation}  
whereas the Frank vector $\Omega^{\star (i)}_k$ will remain invariant together with its scalar product with the Burgers vector. So, if in $2D$ the Frank vector of a given defect line $L^{(i)}$ does not vanish 
($\Omega^{\star (i)}_z\neq 0$), an appropriate change of $x_0$ can always generate arbitrary values of the edge Burgers vector components $B^{\star (i)}_\alpha$ for this particular line.

Suppose now $x_0$ can be selected in such a way that all the Burgers vectors $B^{\star (i)}_k$ vanish while the Frank vectors do not ($\Om^{\star (i)}_z\neq 0$). Then by theorem \ref{mainresultglob}, the incompatibility $\eta^\star_k$ does not vanish, and hence by (\ref{sup2}) the Burgers tensor is not curl-free and cannot be the distributional gradient of a single-valued Burgers field $b^\star_k$ in the absence of translational defects for this particular reference point $x_0$.

\subsection{Completed Frank and Burgers tensors}
\label{sec:compl}

In order to resolve the problem posed in \S\ref{multi_inc}\ref{pospbl} , the tensors $\overline\partial_j\omega^\star_k$ and $\overline\partial_l b^\star_k$ are completed by appropriate concentrated effects within the defect lines, without however modifying their relationship with the multiple-valued Burgers and rotation fields on $\Om_\Lr$. These tensors are called the completed Frank and Burgers tensors.

\begin{Def}\label{defmeso}
\begin{eqnarray}
\hspace{-30pt}\mbox{\scriptsize{COMPLETED FRANK TENSOR }}\hspace{34pt}\eth_j\omega_k^\star&:=&\overline\partial_j\omega^\star_k-\kappa^\star_{kj},\label{disclindens}\\
\hspace{-25pt}\mbox{\scriptsize{COMPLETED BURGERS TENSOR}}\hspace{28pt}\eth_j b^\star_k&:=&\mathcal{E}^\star_{kj}+\epsilon_{kpq}(x_p-x_{0p})\eth_j\omega^\star_q.\label{dislocdens}
\end{eqnarray}
\end{Def}

The following theorems justify the introduction of the completed Frank and Burgers tensors.
\begin{theorem}\label{nouveautenseur1}
In $2D$, the $2$nd-order-tensor distribution $\eth_j\omega_k^\star$ verifies the relation:
\begin{eqnarray}
&&\hspace{-125pt}\mbox{\scriptsize{DISCLINATION DENSITY}}\hspace{65pt} \Theta^\star_{ik}=\epsilon_{ilj}\partial_l\eth_j\omega_k^\star.\label{disclindens2}
\end{eqnarray}
\end{theorem}
{\bf Proof.} 
This statement is a mere consequence of  (\ref{kroner1}) and the relation (\ref{sup1}) expressing incompatibility as the curl of the Frank tensor.{\hfill $\square$}

\begin{theorem}\label{nouveautenseur2}
In $2D$, the $2$nd-order-tensor distribution $\eth_j b^\star_k$ verifies the relation:
\begin{eqnarray}
&&\hspace{-145pt}\mbox{\scriptsize{DISLOCATION DENSITY}}\hspace{48pt} \Lambda^\star_{ik}=\epsilon_{ilj}\partial_l\eth_j b^\star_k.\label{dislocdens2c}
\end{eqnarray}
\end{theorem}
{\bf Proof.}  This statement directly follows from (\ref{sup2}), (\ref{disclindens}) and (\ref{dislocdens}).{\hfill $\square$}
\nl

From the above analysis, it results that the curls of the completed Frank or Burgers tensors
vanish in $\Om$ in the absence of rotational or translational line defects and that in these respective cases these tensors are equal to the gradients of existing single-valued rotation or Burgers vector fields. It is worth noting that the same concentrated contortion term $\kappa^\star_{kj}$ is substracted from the Frank tensor in (\ref{disclindens}) and (\ref{dislocdens}) in order to provide the completed Frank and Burgers tensors.

\subsection{Integral relations and Stokes' theorem}

Returning to $2D$ tensor notations, the following result restates the main theorem \ref{mainresultglob} or \ref{kronerident} in terms of the completed Frank and Burgers tensors (\ref{disclindens}) and (\ref{dislocdens}) and the associated multivalued rotation and Burgers vector fields.
\begin{theorem}\label{kronerident2}
Under assumptions \ref{asstrain} and \ref{assfranktens}, the mesoscopic strain incompatibility for a countable set $\Lr$ of rectilinear dislocations writes as
\begin{eqnarray}
\eta^\star_k&=&\delta_{kz}\epsilon_{\alpha\beta}\partial_\alpha\eth_\beta\omega_k^\star+\epsilon_{\alpha\beta}\partial_\alpha\left(\delta_{kz}\epsilon_{\gamma\tau}\partial_\gamma\eth_\tau b^\star_\beta-\frac{1}{2}\delta_{k\beta}\epsilon_{\gamma\tau}\partial_\gamma\eth_\tau b^\star_z\right),\nonumber
 \end{eqnarray}
with in $\OmLr$, the index-$1$ multivalued fields $\omega^\star_k$ and $b^\star_k$ given by
\begin{eqnarray}
 \omega^\star_k(x)=\omega^\star_{k 0}+\int_{x_0}^x\eth_\beta\omega_k^\star\ dx_\beta\quad\mbox{and}\quad
b^\star_k(x)=b^\star_{k 0}+\int_{x_0}^x\eth_\beta b_k^\star\ dx_\beta.\nonumber
\end{eqnarray} 
\end{theorem}
{\bf Proof.}  This proposition directly results from the introduction of (\ref{disclindens}) and (\ref{dislocdens}) in the main theorem.{\hfill $\square$}
\nl

Line integration of the completed Frank and Burgers tensors in $\OmLr$ therefore provides the rotation and Burgers fields. When this integration is carried out on a loop enclosing a corresponding $2D$ area, the dislocation and disclination densities can themselves be integrated on this area.

\begin{theorem}[Stokes' theorem for the completed defect tensors]
Consider under assumptions  \ref{asstrain} and \ref{assfranktens} a countable set of dislocations and/or disclinations and a $2D$ open set $S\subset\Om_{z_0}$ perpendicular to the defect lines and bounded by the counter clockwise-oriented Jordan curve $C\subset\OmLr$ which encloses once the defect subset $\Lr_C:=\{L^{(i)},\ L^{(i)}\cap S\neq\emptyset,\ i\in\mathcal{I}\}$. Then the following equalities hold:
\begin{eqnarray}
\int_C\eth_\beta \omega_k^\star dx_\beta&=&\int_S\epsilon_{\alpha\beta}\partial_\alpha\eth_\beta \omega_k^\star dS=\int_S\Theta^\star_k dS=\sum_{L^{(i)}\in\mathcal{L}_C}\Om^{\star (i)}_z\delta_{zk},\label{dislocdens3}\\
\int_C\eth_\beta b_k^\star dx_\beta&=&\int_S\epsilon_{\alpha\beta}\partial_\alpha\eth_\beta b_k^\star dS=\int_S\Lambda^\star_k dS=\sum_{L^{(i)}\in\mathcal{L}_C}B^{\star (i)}_k.\label{dislocdens3a}
\end{eqnarray}
\end{theorem}
{\bf Proof.}  Since (\ref{bounded}) results from the assumptions, the dislocation and disclination densities are Radon measures on $\Om_{z_0}$ and hence can be integrated on $S$. Then (\ref{dislocdens3}) and (\ref{dislocdens3a}) directly result from (\ref{disclindens2}), (\ref {disclindens1}), (\ref{dislocdens2b}) and (\ref{dislocdens2}).{\hfill $\square$}

\begin{Rem}
The vector $\eth_\beta\omega_z^\star$ does not verify Stokes' theorem, neither in the classical sense, since $\epsilon_{\alpha\beta}\partial_\alpha\eth_\beta\omega_z^\star$ is singular at $\hat x^{(i)}$, nor in a measure theoretical sense, since 
$\epsilon_{\alpha\beta}\partial_\alpha\eth_\beta\omega_z^\star$ is not a measure but a first-order distribution. The same remark can be made about the Burgers tensor. Nonetheless, as often observed in the literature, even in an inappropriate context the formal use of Stokes' theorem here gives a correct final result.
\end{Rem}

\section{Concluding remarks}\label{concl}
This paper is part of a work devoted to the development of a mathematical theory to analyse dislocated single crystals at the meso-scale by combining distributions with multiple-valued kinematic fields. The distributions are concentrated along the defect lines which in turn form the branching lines of the multivalued fields. From this analysis, a basic theorem relating the incompatibility tensor to the Burgers and Frank vectors of the dislocations and disclinations has been established in the case of countably many defect lines, under precise hypotheses on the distributional elastic strain gradient (via the Frank tensor). Quite surprisingly the sums of the norms of the Burgers and Frank vectors of the defect lines -- which can be derived from the elastic strain -- are required to be locally bounded to obtain the proof, thereby providing a fundamental defect norm for a further homogenization of the medium properties from meso- to macro-scale. This latter problem is addressed in Van Goethem \& Dupret (2009b). 

Moreover, in addition to the elastic strain, two key objective internal fields (the completed Frank and Burgers tensors) have been identified to represent the medium defective state independently of the selection of the reference configuration. While the curls of these two first-order distributional tensors are precisely the disclination and dislocation densities,  their recursive line integration in the defect-free region provides the multiple valued rotation and displacement fields. 

After homogenization from meso- to macro-scale, no concentrated effects will remain present anymore in the macroscopic model, which will consist of a set of evolution PDE's governing the tensorial defect densities in the framework of elasto- or visco-plasticity (cf. e.g. Kratochvil \& Dillon 1969). More precisely, the thermo-mechanical macroscopic model will govern the homogenized elastic strain and completed Frank and Burgers second-order tensors. Let us also mention that the non-vanishing mesoscopic elastic strain incompatibility will generate a macroscopic plastic strain which cannot be defined independently of the choice of the reference configuration. This property simply shows to be a reminiscence of the mesoscopic displacement and rotation multivaluedness.

\end{document}


\title{A distributional approach to the geometry of $2D$ dislocations at the mesoscale\\\begin{normalsize}\textit{Part A: General theory and Volterra dislocations}                                                                                        \end{normalsize}}
\author{Nicolas Van Goethem$^{1}$\& Fran\c{c}ois Dupret$^{2}$ \\
$^{1}$Centro de Matem\'{a}tica e Aplica\c{c}\~{o}es Fundamentais,\\
Universidade de Lisboa,\\
Av. Prof. Gama Pinto, 1649-003 Lisboa, Portugal\\
$^{2}$CESAME, Universit\'e catholique de Louvain,\\
4 av. G. Lema\^{\i}tre, 1348 Louvain-la-Neuve,\ Belgium}
\maketitle
\date{}
\noindent\small{\textbf{Keywords}: dislocations, single crystals, multi-scale analysis, homogenisation, distribution theory, multivalued functions}

\begin{abstract}
We develop a mathematical theory to represent dislocations and disclinations in single crystals at the mesoscopic scale by considering concentrated effects, governed by the distribution theory, combined with multiple-valued kinematic fields.
Our approach provides a new understanding of the continuum theory of defects as developed by Kr\"oner (1980) and other authors. The fundamental identity relating the incompatibility tensor to the Frank and Burgers vectors is proved in the $2D$ case under appropriate assumptions on the strain curl behaviour in the vicinity of the assumed isolated defect lines. In general our theory provides a rigorous framework for the treatment of crystal line defects at mesoscopic scale. Eventually this work will represent a basis to strengthen the mathematical theory of homogenization from mesoscopic to macroscopic scale.
\end{abstract}

\section{Introduction}
\label{intro}

Dislocations can be considered as the most complex class of defects for several kinds of single crystals (V\"{o}lkl \& M\"{u}ller  1994; Jordan \textit{et al.} 2000) and the development of a relevant and accurate physical model represents a key issue with a view to reducing the dislocation density in the crystal by acting on the temperature field and the solid-liquid interface shape during the growth process (Dupret and Van den Bogaert 1994). 

However the dislocation models available in the literature, such as the model of Alexander and Haasen (1986), are often based on a rather crude extension of models initially developed for polycrystals (as usual metals and ceramics are). In this case, some particular features of single crystals, such as material anisotropy or the existence of preferential glide planes, can be taken into account up to some extent, but the fundamental physics of dislocations in single crystals cannot be captured. In fact, dislocations are lines that either form loops, or end at the single crystal boundary, or join together at some locations, while each dislocation segment has a constant Burgers vector which exhibits additive properties at dislocation junctions.
These properties play a fundamental role in the modelling of line defects in single crystals and induce key conservation laws at the macro-scale (typically defined by the crystal diameter).
On the contrary, no dislocation conservation law exists at the macro-scale for polycrystals since dislocations can abruptly end at grain boundaries inside the medium without any conservation law holding across these interfaces.

Aware of these principles and of the pioneer works of Volterra (1907) and Cosserat (1909), Burgers (1939), Eshelby (1956, 1966), Eshelby, Frank \& Nabarro (1951), Kondo (1952), Nye (1953), and Kr\"oner (1980) among other authors (Bilby 1960; Nabarro 1967; Mura 1987) consider a tensorial density to model dislocations in single crystals at the macro-scale, in order to take into account both the dislocation orientation and the associated Burgers vector (cf. the survey contributions of Kr\"oner 1980, 1990,  Kleinert 1989 and Maugin 2003). However, in these works, the relationship between the macro-scale crystal properties and the basic physics governing the nano-scale (defined by the inter-atomic distance) is not completely justified from a mathematical viewpoint. Therefore, to well define the concept of tensorial dislocation density, we here introduce the meso-scale as defined by some average distance between the dislocations. The laws governing the dislocation behaviour are modelled at the nano-scale, while the meso-scale (defined from the nano-scale by ensemble averaging or by averaging over a representative volume (Kr\"oner 2001)) defines the "dislocated continuous medium", where each dislocation is viewed as a line and the interactions between dislocations can be modelled while the laws of linear elasticity govern the adjacent medium.

The present paper focuses on meso-scale modelling with a further view to clarifying the homogenization process from meso- to macro-scale. This latter issue is addressed in the companion work of Van Goethem \& Dupret (2009$a$). Since dislocations are lines at the meso-scale, concentrated effects must be introduced in the mesoscopic model as governed by the distribution theory (Schwartz 1957). In addition, since integration around the dislocations generates a multiple-valued displacement field with the dislocations as branching lines, multivalued functions must be considered (cf, e.g., Almgren 1986). This combination of distributional effects and multivaluedness is a key feature of the dislocation theory at the meso-scale but unfortunately the difficulties resulting from this mathematical association have not well been addressed so far in the literature (see also Thom 1980). As an example, non-commuting differentiation operators are freely introduced without any justification by Kleinert (1989). Therefore, the principal objective of this paper is to provide a strong mathematical foundation to the meso-scale theory of dislocations, showing how the distribution and geometric measure theories can be correctly used with multiple-valued fields. 

In fact, a key modelling issue arises from the fact that homogenization from meso- to macro-scale has no meaning for multiple-valued fields such as displacement and rotation, since this operation is exclusively allowed for additive (or extensive) fields such as stress, energy density or  heat flux. This observation becomes obvious when homogenization is defined by an ensemble averaging procedure, since multiple-valued fields are mathematically defined as extended functions which cannot be added since their "domains" depend on the defect line locations. This consideration justifies the present analysis. For the sake of generality, disclinations, which represent a second but rarer kind of line defect, with in addition a multiple-valued rotation field, are here considered together with dislocations. 

In the literature the macroscopic dislocation density is classically defined as the curl of the plastic distortion (Head \textit{et al.} 1993; Cermelli \& Gurtin 2001; Koslowski \textit{et al.} 2002), following a postulated distortion decomposition into elastic and plastic parts. However, this decomposition cannot be rigorously justified (contrarily to the strain decomposition) since elastic and plastic rotations cannot be set apart without some hidden arbitrariness. In contrast, the present paper paves the way for a rigourous definition and treatment of the macroscopic dislocation density, as obtained from well-defined mesoscopic fields under precise geometric-measure model assumptions, and from which the distortion decomposition can be obtained together with its relationship with the dislocation density (Van Goethem \& Dupret 2009$a$). 

The present paper is restricted to the mesoscopic $2D$ theory for a set of assumed isolated dislocations and/or disclinations. This theory is extended to the case of countably many dislocations in Van Goethem \& Dupret (2009$b$) where the appropriate mathematical objects and functional spaces are ultimately defined for homogenization to the macro-scale. This latter paper will be referred to as Part B in the sequel. Extension to the dynamic $3D$ case is under investigation. Eventually, the complete link between the mesoscopic and macroscopic behaviours of single crystals with line defects should be derived from these developments. In \S \ref{prelim}, the scaling analysis summarized in this introduction is detailed and the basic concepts used to represent the dislocated continuous medium are introduced. The general mathematical theory is developped in \S \ref{multivalued}, while in \S \ref{distranal2}, the $2D$ distributional theory of the dislocated continuous medium is established in the case of isolated parallel  dislocations/disclinations. Conclusions are drawn in \S \ref{concl}. 

\section{Multiscale analysis of dislocations}\label{prelim}

\subsection{Nano-scale analysis: crystalline lattice}\label{nano}
At the nano-scale the characteristic length is the interatomic distance and the reference body is a perfect lattice. Given a dislocation in the general sense (dislocation and/or disclination), the atomic arrangement at time $t$ generaly differs from the reference arrangement, but however the atom displacements are not uniquely defined (Kleinert 1989). Indeed any atom of the reference configuration can in principle be selected to define the displacement of a given atom of the actual configuration which therefore  is a multivalued discrete mapping. Moreover, in general, the dislocation position cannot be determined precisely at the atomic level since several dislocation locations in the actual crystal can be associated with the same picture of the atom positions. In fact the defect should be understood as located inside a nanoscopic lattice region.

\subsection{Meso-scale analysis: dislocated continuous medium an associated reference configurations}\label{meso}

At the meso-scale the characteristic length is some average distance between two neighbour dislocation lines. This scale is the one on which this paper focuses, in the framework of $2D$ linear elasticity. At time $t$, the body is referred to as ${\mathcal R}^{\star}(t)$ as corresponding a random sample corresponding to a given growth experiment. 

A \textit{reference configuration} $\mathcal{R}^\star_0$ with respect to the actual configuration $\mathcal{R}^\star(t)$ is any selected one-to-one transformation of $\mathcal{R}^\star(t)$. $\mathcal{R}^\star_0$ may be chosen as being the body at any given (past or future) time $t_0$ or in contrast be a fictitious transformation of $\mathcal{R}^\star(t)$, and the displacement and rotation fields on $\mathcal{R}^\star(t)$ ($u^\star_i$ and $\omega^\star_k$) are then defined with respect to the chosen $\mathcal{R}_0^\star$. In the present $\mathcal{R}^\star_0$ will always be defined as stress-free and without dislocations, but its selection will remain arbitrary up to this restriction and hence (and this is a keypoint) the defect governing laws must be invariant with respect to the choice of $\mathcal{R}^\star_0$. 

It will be precised later that displacement and rotation are multivalued fields at the mesoscale, and hence are defined on a set called a Riemann foliation $F$ (and not of $\mathcal{R}^\star_0$). The set $F$ can be univoquely associated to $\mathcal{R}^\star(t)$ if a cut is introduced in the foliation in order to select one particular branch of the displacement and rotation.

In view of multivaluedness and the existence of a family of acceptable reference configurations, the main field of our analysis is the assumed linear elastic strain which is clearly single valued and independent of the choice of $\mathcal{R}^\star_0$. The Burgers vector $B^\star_i$ and Frank vector $\Om^\star_i$ are key invariant quantities related to the jump of the multivalued displacement and rotation fields, as directly derived from the linear strain. Their precise definition will be given in \S \ref{multivalued}.
\nl

At this stage, some definitions and assumptions have to be introduced.
\begin{Not}
 In the following sections, the assumed open domain is denoted by $\Omega$ (in practice but not necessarily $\Omega$ is bounded), the defect line(s) are indicated by $\Lr\subset\Omega$, and $\OmLr$ is the chosen symbol for $\Om\setminus\Lr$, which is also assumed to be open.
\end{Not}

\begin{Def}[$3D$ mesoscopic defect lines]\label{meso3D}
At the meso-scale, a $3D$ set $\Lr$ of dislocations and/or disclinations is defined as a set of isolated rectifiable arcs $L^{(k)},\ k\in\mathcal{I}$, without multiple points except possibly their extremities and on which the linear elastic strain is singular. Here a set of isolated arcs means a set of arcs: (i) whose extremeties form a set of isolated points of $\Om$ in the classical sense and (ii) such that each point $\hat x$ of these arcs except their extremities can be located in a smooth surface $S(\hat x)$ bounded by a loop $C(\hat x)$ and such that $S(\hat x)\setminus\hat x\in\Om_\Lr$.
\end{Def}

\begin{Ass}[$3D$ mesoscopic elastic strain]\label{eta}
Henceforth we will  assume that the linear strain $\mathcal{E}^\star_{mn}$ is a given symmetric $L^1(\Om)$ tensor \footnote{It should be noted that $L^q(\Om)$ with any $1\leq q<2$ would hold as well.} prolonged by $0$ on the dislocation set $\Lr$ and compatible on $\OmLr$. In other words,
the incompatibility tensor, as defined by
\begin{eqnarray}
\eta^\star_{kl}:=\epsilon_{kpm}\epsilon_{lqn}\partial_p\partial_q\mathcal{E}^\star_{mn},\label{eta2}
\end{eqnarray}
where derivation is intended in the distribution sense, is assumed to vanish everywhere on $\OmLr$.
\end{Ass}
Let us now introduce the dislocation and disclination density tensors ($\Lambda^\star_{ij}$ and $\Theta^\star_{ij}$) which are the basic physical tools that will be used to model defect density at the meso-scale. 

\begin{Def}[defect densities]
\begin{eqnarray}
&&\hspace{-38pt}\mbox{\scriptsize{DISCLINATION DENSITY:}}\hspace{35pt}\Theta^\star_{ij}:=\sum_{k\in\mathcal{I}\subset\NN}\Om^{\star (k)}_j\tau_i^{(k)}\delta_{L^{(k)}} (i,j=1\cdots 3),\label{disclindens1}\\
&&\hspace{-38pt}\mbox{\scriptsize{DISLOCATION DENSITY:}}\hspace{38pt}\Lambda^\star_{ij}:=\displaystyle\sum_{k\in\mathcal{I}\subset\NN}B^{\star (k)}_j\tau_i^{(k)}\delta_{L^{(k)}} (i,j=1\cdots 3),\label{dislocdens2}
\label{dislocdens4}
\end{eqnarray}
where symbol $\delta_{L^{(k)}}$ is used to represent the one-dimensional Hausdorff measure \footnote{The reader is referred to Mattila (1995) for details on Hausdorff measures.} density concentrated on the rectifiable arc $L^{(k)}$ with the tangent vector $\tau_i^{(k)}$ defined almost everywhere on $L^{(k)}$, while $\Om^{\star (k)}_j$ and $B^{\star (k)}_j$ denote the Frank and Burgers vectors of $L^{(k)}$, respectively. 
\end{Def}

The present paper and Part B address the $2D$ problem only. Generalization of our theory to the $3D$ case will be considered in further publications.
\begin{Def}[$2D$ mesoscopic defect lines]\label{lines2D}
At the meso-scale, a $2D$ set $\Lr$ of dislocations and/or disclinations is defined as a set of isolated parallel lines $L^{(i)},\ i\in\mathcal{I}$, on which the linear elastic strain is singular. In the sequel, these lines will be assumed as parallel to the $z$-axis. 
\end{Def}

More complex sets of $2D$ defect lines are considered in Part B.
\begin{Rem}
 The term $2D$ here refers to the structure of the countable union of points, denoted by $\lr_0$, located at the intersection between $\Lr$ and the $z=z_0$-plane. In this context, the strain is said $2D$ if it solely depends upon the coordinates $x_\alpha\in\Om_{z_0}$ ($\alpha=1,2$). In that case, the displacement and rotation fields will generaly depend on the three space variables.
\end{Rem}

\begin{Not}[$2D$ defect densities and incompatibility]
In $2D$, the vectors $\eta^\star_k,\Theta^\star_k$ and $\Lambda^\star_k$ will denote the tensor components $\eta^\star_{zk},\Theta^\star_{zk}$ and $\Lambda^\star_{zk}$. Greek indices will be used to denote the values $1,2$ (instead of the Latin indices used in $3D$ to denote the values $1,2$ or $3$). Moreover, $\epsilon_{\alpha\beta}$ will denote the permutation symbol $\epsilon_{z\alpha\beta}$.
\end{Not}
The disclination and dislocation density tensors $\Theta^\star_k$ and $\Lambda^\star_k$
will be shown in this paper to be related by a fundamental distributional relation to the strain incompatibility $\eta_k^\star$. In fact, under suitable assumptions on the strain curl (the so-called Frank tensor), the following theorem will be proved in the $2D$ linear elastic case. 
\nl\nl
\textbf{Main theorem:}
\nl
\textbf{incompatibility decomposition for $2D$ isolated defect lines.}
\nl\nl
\textit{The mesoscopic strain incompatibility for a set of isolated parallel rectilinear dislocations $\Lr$ writes as
\begin{eqnarray}
\eta^\star_k=\Theta^\star_k+\epsilon_{\alpha\beta}\partial_\alpha\kappa_{k\beta}^\star,\label{KR1}
\end{eqnarray}
where $\kappa_{k\beta}^\star$ denotes the contortion tensor,
\begin{equation}
\kappa_{k\beta}^\star=\delta_{kz}\alpha^\star_\beta-\frac{1}{2}\alpha^\star_z\delta_{k\beta},\label{KR2} \end{equation} 
with $\alpha_k^\star$ standing for an auxiliary defect density vector,
\begin{equation}
\alpha^\star_k:=\Lambda^\star_k-\delta_{k\alpha}\epsilon_{\alpha\beta}\Theta^\star_z(x_\beta-x_{0\beta}),\label{KR3}
\end{equation} 
and where $x_0$ is a selected reference point in $\Om$.}
\nl

The latter result appears in Kr\"{o}ner's work (1981) under assumptions which are not compatible with our approach. In fact, in his work this result follows in a straightforward manner from an ``elastic-plastic'' displacement gradient (or distortion) decomposition postulate, which itself requires the selection of a particular reference configuration and does not properly handle the intrinsic multivaluedness of the mesoscopic problem. Moreover, in our result the link between the defect densities and the Frank and Burgers vectors is clearly made, and precise assumptions on the strain field and the admissible defect structures are provided in order to validate the result. 

The above theorem will be generalized in Part B to the case of a countable union of parallel rectilinear dislocations. Eventually, the required ``single-valued'' distributional fields will be defined in the appropriate functional spaces for their homogenization to the macro-scale.

\section{Multiple-valued fields and line invariants in the $3D$ case; distributions as a modelling tool at the meso-scale}\label{multivalued}

\begin{Not}\label{notube}
In the following sections, $\hat x$ or $\hat x_i$ will denote a generic point of the defect line(s), $x$ or $x_i$ a generic point of $\Om_\Lr$, and $x_0$ or $x_{0i}$ a given fixed reference point of $\Om_\Lr$. When $x$ and $\hat x$ are used together, $\hat x$ denotes the projection of $x$ onto a given defect line in an appropriate sense and $\hat{\nu}_i:=\nu_i(\hat x,x)$ is the unit vector joining $\hat x$ to $x$. The symbol $\odot_\epsilon$ is intended for a set of diameter $2\epsilon$ enclosing the region $\Lr$. More precisely, $\odot_\epsilon$ is defined as the intersection with $\Om$ of the union of all closed spheres of radius $\epsilon$ centred on $\Lr$: 
\begin{eqnarray}
\displaystyle\odot_\epsilon:=\Omega\cap\bigcup_{\hat x \in \Lr}B[\hat x,\epsilon].\nonumber
\end{eqnarray} 
\end{Not}

If $\Lr$ consists of an single line $L$, $\odot_\epsilon$ is a tube of radius $\epsilon$ enclosing $L$. 

\begin{Not}\label{notube2}
In the sequel, considering a surface $S$ of $\Om$ crossed by a dislocation $L$ at $\hat x$ and bounded by the curve $C$, symbols $dC$, $dL$, and $dS$ will denote the $1D$ Hausdorff measures on $C$ and $L$, and the $2D$ Hausdorff measure on $S$, respectively, with $\tau_j$ standing for the unit tangent vector to $L$ at $\hat x$ (when it exists). In some cases (having fractal curves in mind) the symbols $dx_k$ and $dS_i:=\epsilon_{ijk}dx^{(1)}_jdx^{(2)}_k$ will stand for infinitesimal vectors oriented along $C$ and normal to $S$, respectively, with in addition $dC_l(x):=\epsilon_{lmn}dx_m\tau_n$ denoting an infinitesimal vector normal to $C$ when $\tau_n=\tau_n(\hat x)$ exists.
\end{Not}

In the present section, the strain is assumed to satisfy assumption \ref{eta} and to be smooth away from $\Lr$.

\subsection{Distributional analysis of $3D$ multiple-valued fields}\label{distranal}

In general, a multivalued function from $\Om_\Lr$ to $\RR^N$ is defined as consisting of a pair of single-valued mappings with appropriate properties:\[F\rightarrow\Om_\Lr\quad\mbox{and}\quad\ F\rightarrow \RR^N,\] where $F$ is the associated Riemann foliation (Almgren 
1986). In the present case of meso-scale elasticity, we will limit ourselves to multivalued functions obtained by recursive line integration of single-valued mappings defined on $\Om_\Lr$. Reducing these multiple line integrals to simple line integrals, the Riemann foliation shows to be the set of equivalence path classes inside $\Om_\Lr$ from a given $x_0 \in \Om_\Lr$ with homotopy as equivalence relationship. Accordingly, a multivalued  function will be called of index $n$ on $\Om_\Lr$ if its $n$-th differential is single-valued on $\Om_\Lr$. No other kinds of multifunctions are considered in this work, whether $\Lr$ is a single line $L$ or a more complex set of defect lines (with possible branchings, etc.).

\begin{Not}
The symbol $\partial^{(s)}_j$ is used for partial derivation of a single- or multiple-valued function whose domain is restricted to $\Om_\Lr$. Locally around $x\in\Om_\Lr$, for smooth functions, the meanings of $\partial^{(s)}_j$ and the classical $\partial_j$ are the same, whereas on the entire $\Om$ the partial derivation operator $\partial_j$ only applies to single-valued fields and must be understood in the distributive sense. A defect-free subset $U$ of $\Om$ is an open set such that $U \cap\Lr=\emptyset$, in such a way that $\partial^{(s)}_j$ and $\partial_j$ coincide on $U$ for every single- or multiple-valued  index-$1$ function.
\end{Not}

In the following essential definition generalizing the concept of rotation gradient to dislocated media, the strain is considered as a distribution on $\Omega$.

\begin{Def}[Frank tensor]\label{franktens}
The Frank tensor $\overline\partial_m\omega_k^\star$ is defined as the following distribution on $\Om$:
\begin{eqnarray}
\overline\partial_m\omega_k^\star:=\epsilon_{kpq}\partial_p\mathcal{E}_{qm}^\star,\label{delta_m_a}
\end{eqnarray}
in such a way that
\begin{eqnarray}
<\overline\partial_m\omega_k^\star,\varphi>:=-\int_\Om\epsilon_{kpq}\mathcal{E}_{qm}^\star\partial_p\varphi dV,\label{delta_m}
\end{eqnarray}
with $\varphi$ a smooth test-function with compact support in $\Omega$.
\end{Def}

In fact, in the vicinity of a defect line the tensorial distribution $\overline\partial_m\omega_k^\star$ is the finite part of an integral when acting against test-functions. Indeed, since $\partial_p\mathcal{E}_{qm}^\star$ might be non $L^1(\Om)$-integrable in view of its possibly too strong singularity near the defect lines, instead of being directly calculated as an integral, $<\epsilon_{kpq}\partial_p\mathcal{E}_{qm}^\star,\varphi>$ must be calculated on $\Om$ as the limit
\begin{eqnarray}
\lim_{\epsilon\to 0}\left(\int_{\Om\setminus\odot_\epsilon}\epsilon_{kpq}\partial_p\mathcal{E}_{qm}^\star\varphi dV
+\int_{\partial\odot_\epsilon\cap\Om}\epsilon_{kpq}\mathcal{E}_{qm}^\star\varphi dS_p\right),\label{FP1}
\end{eqnarray}
where the second term inside the parenthesis is precisely added in order to
achieve convergence. One readily sees after integration by parts that (\ref{FP1}) is equal to  (\ref{delta_m}) provided $\displaystyle \lim_{\epsilon\to 0}\Om\setminus \odot_\epsilon=\Om_\Lr$ (which is a general hypothesis limiting the acceptable defect lines and certainly holds true for the lines satisfying definition \ref{meso3D}).

Considering the possibly index-$1$  multivalued rotation vector $\omega^\star_k$, it should be observed from definition \ref{franktens} that $\overline\partial_m\omega_k^\star=\partial^{(s)}_m\omega^\star_k$ on $\Om_\Lr$ as a consequence of the classical relationship between infinitesimal rotation and deformation derivatives. However, $\overline\partial_m\omega_k^\star$ is defined by (\ref{delta_m_a}) as a distribution and therefore concentrated effects on $\Lr$ and its infinitesimal vicinity have to be added to $\partial^{(s)}_m\omega^\star_k$, justifying the use of the symbol $\overline\partial_m\omega^\star_k$ instead of $\partial_m\omega^\star_k$ without giving to $\overline\partial_m$ the meaning of an exact derivation operator. In particular, it may be observed that the identical vanishing of $\partial^{(s)}_m\omega^\star_k$  on $\Om_\Lr$ does not necessarily imply that the distribution $\overline\partial_m\omega_k^\star$ vanishes as well. In fact from  (\ref{FP1}), it can be shown in that case that
\begin{eqnarray}
<\overline\partial_m\omega_k^\star,\varphi>=\lim_{\epsilon\to 0}\int_{\partial\odot_\epsilon\cap\Om}\epsilon_{kpq}\mathcal{E}_{qm}^\star\varphi dS_p=-\int_\Om\epsilon_{kpq}\mathcal{E}_{qm}^\star\partial_p\varphi dV,
\end{eqnarray} 
which is generally non-vanishing. Finally, as soon as the definition of the tensor distribution $\overline\partial_m\omega_k^\star$ is given, so are the distributional derivatives of $\overline\partial_m\omega_k^\star$:
\begin{eqnarray}
<\partial_l\overline\partial_m\omega_k^\star,\varphi>=-<\overline\partial_m\omega_k^\star,\partial_l\varphi>
=\int_\Om\epsilon_{kpn}\mathcal{E}_{mn}^{\star}\partial_p\partial_l\varphi dV.\label{FP2}
\end{eqnarray}

\subsection{$3D$ rotation and displacement vectors}\label{rotdispl}
The rotation vector is defined from the linear strain together with the rotation at a given point $x_0$. From this construction follows an invariance property of $\omega^\star_k$ as a multifunction (recalling that  multivaluedness takes its origin from the existence of defect lines which render the strain incompatible on the entire $\Om$).\newline

Starting from the distributive definition \ref{franktens} of $\overline\partial_m\omega_k^\star$, the form $\overline\partial_m\omega^\star_k$ $d\xi_m$ is integrated along a regular parametric curve $\Gamma \subset \Om_\Lr$ with endpoints $x_0, x\in \Om_\Lr$. For selected $x_0$ and $\omega^\star_{0k}$, the multivalued rotation vector is defined as \footnote{For a non-smooth strain, integration is to be understood in the distribution sense.}
\begin{eqnarray}
\omega^\star_k=\omega^\star_k(\#\Gamma;\omega^\star_0)=\omega^\star_{0k}+\int_\Gamma\overline\partial_m\omega^{\star}_kd\xi_m,\nonumber
\end{eqnarray}
where $\#\Gamma$ is the equivalence class of all regular curves homotopic to $\Gamma$ in $\Om_\Lr$. Indeed, from strain compatibility in $\Om_\Lr$, i.e. from relation (\ref{eta2}), it is clear that $\omega^\star_k$ is a function of $\#\Gamma$  only. Consider now a regular parametric loop $C$ (in case $C$ is a planar loop, it is a Jordan curve) and the equivalence class $\#C$ of all regular loops homotopic to $C$ in $\Om_\Lr$. Here, the extremity points play no role anymore and two loops are equivalent if and only if they can be continuously transformed into each other in $\Om_\Lr$. The jump of the rotation vector $\omega^\star_k$ along $\#C$ depends on $\#C$ only and is calculated as \footnote{We note that $C$ could be non rectifiable, i.e. of infinite length. Integrals on fractal curves and the related Stokes' and Gauss-Green's theorems are analysed by Harrison \& Norton (1992), where it is shown by the $\mathcal{C}^\infty$ smoothness of the differential form $\overline\partial_m \omega^\star_k dx_m$ on $\Om_\Lr$ that  (\ref{GG}) still holds even when the Hausdorff dimension of $C$ is higher than $1$.}
\begin{eqnarray}
[\omega^\star_k]=[\omega^\star_k](\#C)=\int_C \overline\partial_m \omega^\star_k d\xi_m.\label{GG}
\end{eqnarray}

The following developments address the displacement field  multivaluedness as a mere consequence of strain incompatibility. The procedure defining the displacement vector from the rotation vector by means of line integrals is classical in linear elasticity. The following tensor plays in the construction of the displacement field a role analogous to $\overline\partial_m\omega^\star_k$ in the construction of the rotation field.

\begin{Def}[Burgers tensor]\label{burgtens}
For a selected reference point $x_0\in\Om_\Lr$, the Burgers tensor is defined on the entire domain $\Om$ as the distribution
\begin{eqnarray}
\displaystyle\overline\partial_l b^\star_k(x;x_0):=\mathcal{E}^\star_{kl}(x)+\epsilon_{kpq}(x_p-x_{0p})\overline\partial_l\omega^\star_q(x).\label{delta_lb_i}
\end{eqnarray}
\end{Def}

The Burgers tensor can be integrated in the same way as the Frank tensor along any parametric curve $\Gamma$, providing for selected $\omega^\star_{0k}$ and $u^\star_{0k}$ the index $2$-multivalued displacement vector $u^\star_k$:
\begin{eqnarray}
u_k^\star=u_k^\star(x,\#\Gamma;\omega^\star_0,u^\star_0)=u^\star_{0k}+\epsilon_{klm}\omega_l^\star(\#\Gamma;\omega^\star_0)(x_m-x_{0m})+\int_\Gamma\overline\partial_l b^\star_k(\xi)d\xi_l,\nonumber
\end{eqnarray}
which is a function of $x$ and $\#\Gamma$ only (this following from (\ref{eta2}) and (\ref{delta_lb_i})). It may be observed that $\overline\partial_l b^\star_k$ and the vector
\begin{equation}
b^\star_k=b^\star_k(\#\Gamma;u^\star_0)=u^\star_k-\epsilon_{klm}\omega_l^\star(x_m-x_{0m}) \label{burgers}
\end{equation} 
are related in the same way as $\overline\partial_m\omega^\star_k$ and $\omega^\star_k$, including the fact that $\overline\partial_l b^\star_k=\partial_l^{(s)} b^\star_k$ on $\Om_\Lr$. The jumps of $b^\star_k$ along $\#C$ and of $u^\star_k$ at $x$ along $\#C$ (which depend on $\#C$ only) are calculated as
\begin{eqnarray}
[b^\star_k](\#C;x_0)=[u_k^\star](x;\#C;x_0)-\epsilon_{klm}[\omega^\star_l](\#C)(x_m-x_{0m})=\int_C \overline\partial_lb^\star_k d\xi_l.
\end{eqnarray}

Let us now focus on the case of a given isolated defect line $L^{(i)}, \ i\in\mathcal{I}$. The jump $[\omega^\star_k]$ of the rotation vector $\omega^\star_k$ around $L^{(i)}$ is defined as the jump of $\omega^\star_k$ along $\#C(\hat x)$, with $\hat x$ a point of $L^{(i)}$ and $C(\hat x)$ a loop enclosing once the defect line $L^{(i)}$ and no other defect line as specified in definition \ref{meso3D}. It turns out that this jump is the same for any $\hat x$ and suitable $C(\hat x)$. Similarly, the jump $[b^\star_k]$ of the vector $b^\star_k$ around $L^{(i)}$ is defined as the jump of $b^\star_k$ along $\#C(\hat x)$ and is also the same for any $\hat x$ and suitable $C(\hat x)$, given $x_0$. In fact, the following result is well-known (Kleinert 1989).

\begin{theorem}[Weingarten's theorem]\label{Wein1}
The rotation vector $\omega_k^{\star}$ is an index-$1$ multifunction on $\Om_\Lr$ whose jump  $\Om^{\star}_k:=[\omega_k^{\star}]$ around the isolated defect line $L^{(i)}, \ i\in\mathcal{I}$, is an invariant of this line. Moreover, for a given $x_0$, the vector $b_k^\star$ is a multifunction of index $1$ on $\Om_\Lr$ whose jump  $B^\star_k:=[b_k^\star]$ around $L^{(i)}$ is an invariant of this line.
\end{theorem}

From this result, the Frank and Burgers vectors are defined as invariants of $L^{(i)}$.
\begin{Def}[Frank and Burgers vectors]\label{Burgers}
The Frank vector of an isolated defect line $L^{(i)}, \ i\in\mathcal{I}$, is the invariant
\begin{eqnarray}
\Om^\star_k:=[\omega^\star_k],\label{frank}
\end{eqnarray}
while for a given reference point $x_0$ its Burgers vector is the invariant
\begin{eqnarray}
B^\star_k:=[b^\star_k]=[u_k^\star](x)-\epsilon_{klm}\Om^\star_l(x_m-x_{0m}).\label{burgers}
\end{eqnarray}
\end{Def}

A defect line with non-vanishing Frank vector is called a disclination while a defect line with non-vanishing Burgers vector is called a dislocation. Clearly a disclination should always be considered as a dislocation by appropriate choice of $x_0$ while the reverse statement is false since $\Om^\star_k$ might vanish. In fact, two distinct reference points $x_0$ and $x'_0$ define two distinct Burgers vectors, obeying the relation $B^\star_k-B'^\star_k=\epsilon_{klm}(x_{0m}-x'_{0m})\Om^\star_l$ (noting that $B^\star_k\Om_k^\star$ is an invariant independent of the arbitrary choice of $x_0$). Therefore, for a non-zero Frank vector, the vanishing of the Burgers vector depends on the arbitrary choice of $x_0$. This is why in the present paper, the word "dislocation" means in the general sense a dislocation and/or a disclination. A pure dislocation is a dislocation with vanishing Frank vector.
\begin{Rem}
 It should be emphasized that the assumption of isolated defect lines is required to construct appropriate enclosing loops in order to define their Frank and Burgers vectors. In Part B this assumption will be removed for a countable set of parallel defect lines under appropriate assumptions on the Frank tensor.
\end{Rem}

In general, every defect line will contribute to the rotation and displacement multivaluedness, and hence these latter fields are defined over $\Om_\Lr$ and do not share the structure of a vector space. In other words, the displacement and rotation fields cannot be added since their domains depend on the defect line locations.

Therefore, besides the strain field which is the seminal ingredient of the present theory, the Burgers and Frank tensors appear as fundamental quantities able to characterize the amount of defects on each single line or in the whole dislocated crystal. Together with the geometry of the defect set, these vectors provide the key defect measures called the dislocation and disclination density tensors (which now belong to a vector space). Accordingly, the following well-known result can be readily shown and is fundamental in the framework of our investigations since it implies conservation laws at the meso- and macro-scales.

\begin{theorem}[conservation laws]
Single disclination and dislocation lines are always closed or end at the boundary of $\Om$. Moreover, in all cases,
\begin{eqnarray}
\partial_i\Theta_{ij}^\star=\partial_i\Lambda_{ij}^\star=0.\nonumber
\end{eqnarray}
\end{theorem}

\section{Distributional analysis of incompatibility for a single rectilinear dislocation}\label{distranal2}
\subsection{The $2D$ model for rectilinear dislocations}\label{2Dsection}

$2D$ elasticity means that the strain $\mathcal{E}^\star_{ij}$ is independent of the "vertical" coordinate $z$. However this assumption introduces no restriction on the dependence of the multiple-valued displacement and rotation fields upon $z$. 
\begin{Not}\label{planarnot}
In this \S \ref{distranal2}\ref{2Dsection}, the single defect line $L$ is assumed to be located along the $z$-axis.
The two planar coordinates will be denoted by $(x,y)$ or $x_\alpha$. The projection of $x=(x_\alpha,z)$ on $L$ is $\hat x=(0,0,z)$ and $(\nu_\alpha, 0)$ stands for the unit vector from $\hat x$ to $x$. Symbols $(e_x,e_y,e_z)$ or $(e_\alpha,e_z)$ denote the Cartesian base vectors, while $(e_r,e_\theta,e_z)$ denote the local cylindrical base vectors. For a planar curve $C$, the notation $dC_\alpha(x)=\epsilon_{\alpha\beta}dx_\beta$ is used for an infinitesimal vector parallel to the curve normal.
\end{Not}

Let us observe that many fields are singular at the origin and that $\Om_L$ is in fact the domain where the laws of linear elasticity apply. Moreover, the strain can  be decomposed into three tensors:
\begin{eqnarray}
\mathcal{E}^\star_{ij}=\underbrace{\delta_{\alpha i}\delta_{\beta j}\mathcal{E}^\star_{\alpha\beta}}_{planar\ strain}+\underbrace{\left(\delta_{i z}\delta_{j \gamma}\mathcal{E}^\star_{\gamma z}+\delta_{j z}\delta_{i \gamma}\mathcal{E}^\star_{\gamma z}\right)}_{3D\ shear}\underbrace{+\delta_{i z}\delta_{j z}\mathcal{E}^\star_{zz}.}_{pure\ vertical\ compression/dilation}\nonumber
\end{eqnarray}

\begin{lemma}[$2D$ compatibility]\label{complane}
In $\Om_L$, from $2D$ strain compatibility, there are real numbers $K, a_\alpha$ and $b$ such that
\begin{eqnarray}
\left\{\begin{array}{lll}\epsilon_{\alpha\gamma}\epsilon_{\beta\delta}\partial_\alpha\partial_\beta\mathcal{E}_{\gamma\delta}^\star=0, \\
\epsilon_{\alpha\beta}\partial_\alpha\mathcal{E}_{\beta z}^\star=K, \\ \mathcal{E}_{zz}^\star=a_\alpha x_\alpha+b.\end{array}\right.\label{eqcomplane}
\end{eqnarray}
\end{lemma}
This lemma is easily proved from assumption \ref{eta}.{\hfill $\square$}
\begin{Rem}
The present theory does not make use of the linear elasticity constitutive laws and the momentum and energy conservation laws, since in the framework of Continuum Mechanics arbitrary body forces and heat supply can always be applied to the medium. Moreover, the sum of these two body contributions and the unsteady terms governing the medium dynamics can generally be nonsmooth, and hence the stress and heat flux derivatives have to be treated as mathematical distributions thereby providing a physical justification to our approach. 
\end{Rem}

The remaining of this section will be devoted to present the three classical examples of $2D$ line-defects for which the medium is assumed to be steady, body force free and isothermal (detail is given in Van Goethem 2007). 

\begin{itemize}
\item \textit{Pure screw dislocation.} The displacement and rotation vectors write as
\begin{eqnarray}
u^\star_ie_i=\frac{B^\star_z\theta}{2\pi}e_z\quad\mbox{and}\quad \omega^\star_ie_i=\frac{1}{2}\nabla\times u^\star_ie_i=\frac{B^\star_z}{4\pi r}e_r,\label{disp_screw}
\end{eqnarray}
in such a way that the jump $[\omega^\star_i]$ vanishes identically, while the Cartesian strain is divergence-free on $\Om$ and writes as
\begin{eqnarray}
[\mathcal{E}^\star_{ij}]=\frac{-B^\star_z}{4\pi r^2}\left[ \begin{array}{ccc} 0& 0 & y\\ 0 & 0 & -x \\ y & -x & 0
\end{array} \right].\label{strain_screw}
\end{eqnarray}
Moreover, inside $\Om_L$, the Frank tensor writes as 

\begin{eqnarray}\label{frank_screw}
[\overline\partial_m\omega^\star_k]=\frac{-B^\star_z}{4\pi r^{2}}\left[\begin{array}{ccc} \cos2\theta & \sin2\theta & 0 \\ \sin2\theta &-\cos2\theta   & 0 \\ 0 & 0 & 0
\end{array}\right].
\end{eqnarray}

\item \textit{Pure edge dislocation.} The displacement vector is
\[
u^\star_i e_i=\frac{-B^\star_y(\log\frac{r}{R}+1)}{2\pi}e_x+\frac{B^\star_y\theta}{2\pi}e_y,
\]
while the rotation $\omega^\star_i$ vanishes together with its jump. The Cartesian strain (which requires additional regular terms to correspond to balanced stresses) writes as
\begin{eqnarray}
[\mathcal{E}^\star_{ij}]=\frac{-B^\star_y}{2\pi r^{2}}\left [ \begin{array}{ccc} x&y& 0 \\ y&-x  & 0  \\ 0 & 0 &0
\end{array} \right ],\label{strain_edge}
\end{eqnarray}
noting that the tensor $\overline\partial_m\omega_k^\star$ vanishes identically inside $\Om_L$.
\item \textit{Wedge disclination.} The rotation vector is \[\omega^\star_ie_i=\frac{\Om^\star_z\theta}{2\pi}e_z,\]
with the multiple-valued planar displacement field given by
\begin{eqnarray}
&u^\star_x-iu^\star_y=\frac{\Om^\star_z}{4\pi}(1-\nu^*)x\ln(\frac{r}{R})-\frac{\Om^\star_z}{8\pi}(1+\nu^*)x-\frac{\Om_z^\star}{2\pi}y\theta&\nonumber\\
&-i\left[\frac{\Om^\star_z}{4\pi}(1-\nu^*)y\ln(\frac{r}{R})-\frac{\Om^\star_z}{8\pi}(1+\nu^*)x+\frac{\Om^\star_z}{2\pi}x\theta\right]\label{u_disclin}
\end{eqnarray}
(where $\nu^*$ stands for the $2D$ Poisson coefficient, to be considered as an arbitrary constant together with $R$) and with a vanishing Burgers vector: \[B^\star_x-iB^\star_y=[u^\star_x]-i[u^\star_y]+\Om^\star_z(y+ix)=0.\]
The Cartesian strain writes as
\begin{eqnarray}
[\mathcal{E}^\star_{ij}]=\frac{\Om_{z}(1-\nu^{\star})}{4\pi}\left [ \begin{array}{ccc} (\log\frac{r}{R}+1)& 0& 0 \\0&(\log\frac{r}{R}+1) & 0  \\ 0 & 0 &0
\end{array} \right ]&&\nonumber
\\-\frac{\Om^\star_z(1+\nu^*)}{8\pi}\left [ \begin{array}{ccc} \cos2\theta& \sin2\theta & 0 \\\sin2\theta &-\cos2\theta & 0  \\ 0 & 0 &0
\end{array} \right],&&\label{strain_disclin}
\end{eqnarray}
and hence
\begin{eqnarray}
[\overline\partial_m\omega^\star_k]=
-\frac{\Om^\star_{z}}{2\pi r}\left[\begin{array}{ccc} 0& 0& \sin\theta \\0&0&-\cos\theta \\ 0 & 0 &0
\end{array}\right].\label{frank_disclin}
\end{eqnarray}
\end{itemize}

\begin{Rem}
The energy density $\mathcal{E}^\star=\frac{1}{2}\sigma_{ij}^\star\mathcal{E}^\star_{ij}$ is not $L^1$-integrable for both kinds of dislocations, while it is finite for the wedge disclination. Therefore, a Hadamard finite part (Schwartz 1957; Estrada \& Kanwal 1989) is needed to represent the compliance at the meso-scale (another approach makes use of strain mollification by a so-called core tensor (Koslowski \textit{et al.} 2002)). This issue, whose solution requires to develop matched asymptotic expansions around the singular line in accordance with the infinitesimal displacement hypothesis, will not be addressed further in the present paper which only focuses on the geometry of dislocations.
\end{Rem}
\begin{Rem}
The above expressions of dislocations and disclinations do not necessarily provide balanced stresses. The present theory is fully independent of any dynamical assumption and only focuses on the geometrical concentrated properties of the defect lines.
\end{Rem}
\subsection{Mesoscopic incompatibility for a single defect line}\label{mesosingle}
For $2D$ problems the incompatibility vector contains all the information provided by the general incompatibility tensor. The latter expresses on the one hand the non-commutative action of the defect line over the second derivatives of the rotation vector and on the other hand is related to concentrated effects of the Frank and Burgers vectors along the defect line.

\begin{Def}[$2D$ incompatibility tensor]\label{incompatibility}
In the $2D$ case, the mesoscopic incompatibility vector is defined by
\begin{eqnarray}
\eta^{\star}_k:=\epsilon_{\alpha\beta}\partial_\alpha\overline\partial_\beta\omega^{\star}_k.\label{eta-om}
\end{eqnarray}
A strain field is compatible if the associated incompatibility vector vanishes.
\end{Def}

As shown in the following sections, concentration effects will be represented by means of first- and second-order distributions.

\begin{Not} \label{newnot1}
Let, with use of notation \ref{planarnot}, $\Om_z:=\{x\in\Om\ \mbox{with a prescribed $z$}\}$ and $\Om_z^0:=\Om_z\setminus L$ while the radius $r=\Vert x-\hat x\Vert$ is the distance from a point $x$ inside $\Om$ to $L$. Then, the $1D$ Hausdorff measure density concentrated on $L$ will be denoted by $\delta_L$.
\end{Not}

In what follows the hypothesis consists in assuming that the strain radial dependence in the vicinity of $L$ is less singular than a critical threshold. This is verified for instance by the wedge disclination whose strain radial behaviour is $O(\ln r)$  \footnote{A function $f(\epsilon)$ is said to be $O\left(g(\epsilon)\right)(\epsilon\to 0^+)$ if there exists $K, \epsilon_0>0$ s.t. $\displaystyle 0<\epsilon<\epsilon_0\Rightarrow |f(\epsilon)|\leq K|g(\epsilon)|$. A function $f(\epsilon)$ is said to be $o\left(g(\epsilon)\right)(\epsilon\to 0^+)$ if $\displaystyle \lim_{\epsilon\to 0^+}\frac{f(\epsilon)}{g(\epsilon)}=0$.} and by the screw and edge dislocations whose strains are $O(r^{-1})$.

For a straight defect line $L$, according to these examples, the hypotheses on the strain and Frank tensors read as follows.

\begin{Ass}[$2D$ strain for line defects]\label{asstrain} 
The strain tensor $\mathcal{E}_{ij}^\star$ is independent of the coordinate $z$, compatible on $\Om_L=\Om\setminus L$ in the sense that conditions (\ref{complane}) hold, smooth on $\Om_L$, and $L^1$-integrable on $\Omega$. 
\end{Ass}

\begin{Ass} [local behaviour]\label{asstrainloc}
The strain tensor $\mathcal{E}_{ij}^\star$ is assumed to be $o(r^{-2})$ $(\epsilon\to 0^+)$ while the Frank tensor is assumed to be $o(r^{-3})(\epsilon\to 0^+)$.
\end{Ass}

The two following lemmas are needed for the proof of our main result for a single isolated defect line.

\begin{lemma}\label{fleurs}
Let $C_\epsilon(\hat x),\ \epsilon>0$, denote a family of $2D$ closed rectifiable curves. Then, in $2D$ elasticity, the Frank tensor and the strain verify the relation
\begin{equation}
\lim_{C_\epsilon(\hat x)\to\hat x}\int_{C_\epsilon(\hat x)}\left(x_\alpha\overline\partial_\beta\omega_\kappa^\star dx_\beta +\epsilon_{\kappa\beta}\mathcal{E}_{\beta z}^\star\right) dx_\alpha=0,\nonumber
\end{equation}
provided the length of $C_\epsilon$ is uniformly bounded and as long as the convergence $C_\epsilon(\hat x)\to\hat x$ is understood in the Hausdorff sense, i.e. in such a way that
\begin{eqnarray}
\max\{\Vert x-\hat x\Vert, x\in C_\epsilon(\hat x)\}\to 0.\nonumber
\end{eqnarray}
\end{lemma}
{\bf Proof.} 
The second compatibility condition of  (\ref{eqcomplane}) is equivalent to
\[\partial_\gamma\mathcal{E}_{\beta z}^\star-\partial_\beta\mathcal{E}_{\gamma z}^\star=K\epsilon_{\gamma\beta},\]
from which, in $2D$ elasticity:
\begin{eqnarray}
\overline\partial_\beta\omega^\star_\kappa:=\epsilon_{\kappa\gamma}\partial_\gamma\mathcal{E}_{\beta z}^\star=\epsilon_{\kappa\gamma}\partial_\beta\mathcal{E}_{\gamma z}^\star-K\delta_{\kappa\beta},\nonumber
\end{eqnarray}
and
\begin{eqnarray}
\left(x_\alpha\overline\partial_\beta\omega_\kappa^\star+\delta_{\alpha\beta}\epsilon_{\kappa\gamma}\mathcal{E}_{\gamma z}^\star\right)=\partial_\beta\Bigl(x_\alpha\epsilon_{\kappa\gamma}
\mathcal{E}_{\gamma z}^\star\Bigr)-x_\alpha K\delta_{\kappa\beta}.\nonumber
\end{eqnarray}
Since, under the assumptions of this lemma,
\begin{eqnarray}
\lim_{C_\epsilon(\hat x)\to \hat x}\int_{C_\epsilon(\hat x)}x_\alpha dx_\kappa=0,\nonumber
\end{eqnarray}
while the strain is a single-valued tensor, the proof is achieved.{\hfill $\square$}

\begin{lemma}\label{racine}
In $2D$ elasticity the planar Frank vector $\Om^\star_\alpha$ vanishes.
\end{lemma}
{\bf Proof.} 
Since
\begin{eqnarray}
\overline\partial_\beta b^\star_\tau=\mathcal{E}^\star_{\beta\tau}+\epsilon_{\tau\gamma}(x_\gamma-x_{0\gamma})\delta_\beta\omega^\star_z
-\epsilon_{\tau\gamma}(z-z_0)\delta_\beta\omega^\star_\gamma,\nonumber
\end{eqnarray}
the planar Burgers vector simply writes as
\begin{eqnarray}
B^\star_\tau=\int_{C}\left(\mathcal{E}^\star_{\beta\tau}+\epsilon_{\tau\gamma}(x_\gamma-x_{0\gamma})\delta_\beta\omega^\star_z\right)dx_\beta
-\epsilon_{\tau\gamma}(z-z_0)\Omega^\star_\gamma,\nonumber
\end{eqnarray}
where $C$ is any planar loop.
By Weingarten's theorems the Burgers vector is a constant while the integrand is independent of $z$, from which the result obviously follows.{\hfill $\square$}

\begin{theorem}[main result for a single defect line]\label{mainresultloc}
Under assumptions \ref{asstrain} and \ref{asstrainloc}, for a dislocation located along the $z$-axis, incompatibility as defined by equation (\ref{eta-om}) is the vectorial first order distribution
\begin{eqnarray}
\eta_k^{\star}=\delta_{kz}\eta_z^{\star}+\delta_{k\kappa}\eta_\kappa^{\star},\label{eta_k}\label{etak}
\end{eqnarray}
with
\begin{eqnarray}
\eta^\star_z&=&\Om^\star_z\delta_L+\epsilon_{\alpha\gamma}\left(B^\star_\gamma-\epsilon_{\beta\gamma}x_{0\beta}\Om^\star_z\right)\partial_\alpha\delta_L,\label{etaz}\\
\eta^\star_\kappa&=&\frac{1}{2}\epsilon_{\kappa\alpha}B^\star_z\partial_\alpha\delta_L\label{etakappa}.
\end{eqnarray}
\end{theorem}
{\bf Proof.} 
For some small enough $\epsilon>0$ and using notations \ref{notube}, a tube $\odot_\epsilon$ can be constructed around $L$ and inside $\Om$. Assuming that the smooth $3D$ test-function $\varphi$ has its compact support containing a part of $L$, $\Om_{\epsilon,z}$ denotes the slice of the open $\Om\setminus\odot_\epsilon$ obtained for a given $\hat x\in L$, i.e.
\begin{equation}
\Om_{\epsilon,z}:=\{x\in\Om_z\quad\mbox{such that}\quad ||x_\alpha||>\epsilon\},\nonumber
\end{equation}
while the boundary circle of $\Om_{\epsilon,z}$ is designated by $C_{\epsilon,z}$.
\newline 
$\blacktriangle$ Let us firstly treat the left-hand side of  (\ref{etak}). From definitions \ref{incompatibility} and \ref{franktens}, and equations (\ref{delta_m_a}), (\ref{delta_m}) and (\ref{FP1}), it follows that
\begin{equation}
<\eta^\star_k,\varphi>=\int_Ldz\lim_{\epsilon\to 0^+}\Pi_k(z,\varphi,\epsilon),\label{FD1}
\end{equation}
where
\begin{equation}
\Pi_k(z,\varphi,\epsilon):=-\int_{\Om_{\epsilon,z}}\epsilon_{\alpha\beta}\overline\partial_\beta\omega^\star_k\partial_\alpha
\varphi dS-\int_{C_{\epsilon,z}}\epsilon_{\alpha\beta}\epsilon_{k\gamma n}\mathcal{E}_{\beta n}^\star\partial_\alpha\varphi dC_\gamma.\label{FD2}
\end{equation}
From notation \ref{planarnot}, the boundedness of $|\partial_\tau\partial_\delta\varphi|$ on $\Om_L$ provides the following Taylor expansions of $\varphi$ and $\partial_\alpha\varphi$ around $\hat x$:
\begin{eqnarray}
\varphi(x)&=&\varphi(\hat x)+r\nu_\alpha\partial_\alpha\varphi(\hat x)+\frac{r^2}{2}\nu_\tau\nu_\delta\partial_\tau
\partial_\delta\varphi\left(\hat x+\gamma_1(x-\hat x)\right),\label{taylor1}\\
\partial_\alpha\varphi(x)&=&\partial_\alpha\varphi(\hat x)+r\nu_\tau\partial_\tau
\partial_\alpha\varphi\left(\hat x+\gamma_2(x-\hat x)\right),\label{taylor2}
\end{eqnarray}
with $0<\gamma_1(x-\hat x),\gamma_2(x-\hat x)\leq 1$.\newline
$\blacktriangle$ Consider the first term of the right-hand side of (\ref{FD2}), noted $\hat\Pi_k$. By virtue of the strain compatibility on $\Om_L$ and Gauss-Green's theorem, this term writes as
\begin{equation}
\hat\Pi_k(z,\varphi,\epsilon):=-\int_{\Om_{\epsilon,z}}\partial_\gamma\left(\epsilon_{\gamma\beta}\overline\partial_\beta\omega^{\star}_k\varphi\right)dS=\int_{C_\epsilon}\epsilon_{\gamma\beta}\overline\partial_\beta\omega^\star_k\varphi dC_\gamma.\nonumber
\end{equation}
Since by notations \ref{planarnot} and \ref{newnot1}, $r\nu_\alpha:=x_\alpha-\hat x_\alpha=x_\alpha$, then equation (\ref{taylor1}) and assumption \ref{asstrainloc} show that, for $\epsilon\to 0^+$,
\begin{eqnarray} 
\hat\Pi_k=\int_{C_{\epsilon,z}}\epsilon_{\gamma\beta}\overline\partial_\beta\omega^\star_k\Bigl(\varphi(\hat x)+x_\alpha\partial_\alpha\varphi(\hat x)\Bigr)dC_\gamma+o(1).\nonumber
\end{eqnarray}
$\blacktriangle$ Consider the second term of the right-hand side of (\ref{FD2}), noted $\Pi_k^\star$. On account of assumption \ref{asstrainloc} and from expansion (\ref{taylor2}), this term may be rewritten as
\begin{eqnarray}
\Pi_k^\star(z,\varphi,\epsilon)&:=&-\int_{C_{\epsilon,z}}\epsilon_{\alpha\beta}\epsilon_{k\gamma n}\mathcal{E}_{\beta n}^\star\partial_\alpha\varphi dC_\gamma\nonumber\\
&=&-\partial_\alpha\varphi(\hat x)\int_{C_{\epsilon,z}}\epsilon_{\alpha\beta}\epsilon_{k\gamma n}\mathcal{E}_{\beta n}^\star dC_\gamma+o(1).\nonumber
\end{eqnarray}
$\blacktriangle$ From Weingarten's theorem and recalling that $dC_\gamma=\epsilon_{\gamma\tau}dx_\tau$, the expression $\Pi_k=\hat{\Pi}_k+\Pi_k^\star$ then writes as
\begin{eqnarray}
\Pi_k=\partial_\alpha\varphi(\hat x)\int_{C_{\epsilon,z}}\left(x_\alpha
\overline\partial_\tau\omega^\star_k-\epsilon_{\alpha\beta}\epsilon_{k\gamma n}\epsilon_{\gamma\tau}\mathcal{E}_{\beta n}^\star\right)dx_\tau&&\nonumber\\+\Om^\star_k\varphi(\hat x)+o(1).&&\label{pik}
\end{eqnarray}
$\blacktriangle$ Consider the first term of the right-hand side of (\ref{pik}), noted $\Pi'_k$, and take $\xi=\gamma$ in the identity
\begin{eqnarray}
\epsilon_{k\xi n}\epsilon_{\gamma\tau}=\delta_{kz}\left(\delta_{\gamma\xi}\delta_{n\tau}-\delta_{n\gamma}\delta_{\tau\xi}\right)-\delta_{nz}\left(\delta_{\gamma\xi}
\delta_{k\tau}-\delta_{k\gamma}\delta_{\tau\xi}\right)\label{identity}
\end{eqnarray}
in such a way that
\begin{eqnarray}
\Pi'_k=
\partial_\alpha\varphi(\hat x)\int_{C_{\epsilon,z}}\left(x_\alpha
\overline\partial_\tau\omega^\star_k-\delta_{kz}\epsilon_{\alpha\beta}
\mathcal{E}_{\beta\tau}^\star+\delta_{k\tau}\epsilon_{\alpha\beta}
\mathcal{E}_{\beta z}^\star\right)dx_\tau.\label{ombre}
\end{eqnarray}
$\blacktriangle$ The cases $k=z$ and $k=\kappa$ are now treated separately.

$\bullet$ When $k=z$, definition \ref{burgtens} shows that
\begin{eqnarray}
\displaystyle\overline\partial_\beta b^\star_\tau:=\mathcal{E}^\star_{\beta\tau}+\epsilon_{\tau\gamma}(x_\gamma-x_{0\gamma})\overline\partial_\beta\omega^\star_z-\epsilon_{\tau\gamma}(z-z_0)\overline\partial_\beta\omega^\star_\gamma\nonumber
\end{eqnarray}
which, after multiplication by $\epsilon_{\tau\alpha}$ and using  (\ref{identity}) with $\tau,\alpha$ and $z$ substituted for $k,\xi$ and $n$, is inserted into  (\ref{ombre}), thence yielding:
\begin{eqnarray}
\Pi'_z=\partial_\alpha\varphi(\hat x)\int_{C_{\epsilon,z}}
\left(\epsilon_{\tau\alpha}\overline\partial_\beta b_\tau^\star+x_{0\alpha}\overline\partial_\beta\omega_z^\star
+(z-z_0)\overline\partial_\beta\omega_\alpha^\star\right)dx_\beta,\label{mousse}
\end{eqnarray}
and consequently, from the definitions of the Frank and Burgers vectors,
\begin{eqnarray}\label{feuilles}
\lim_{\epsilon\to 0^+}\Pi'_z=\ \ll\left\{\epsilon_{\alpha\tau}B_\tau^\star-(z-z_0)\Om_\alpha^\star-x_{0\alpha}\Om_z^\star\right\}\partial_\alpha\delta_0,\varphi_z\gg,
\end{eqnarray}
where $\delta_0$ is the $2D$ Dirac measure located at $0$ and $\varphi_z(x_\alpha):=\varphi(x_\alpha,z)$, while symbol $\ll\cdot,\cdot\gg$ denotes the $2D$ distribution by test-function product.

$\bullet$ When $k=\kappa$, definition \ref{burgtens} shows that
\begin{eqnarray}
\displaystyle\overline\partial_\beta b^\star_z:=\mathcal{E}^\star_{\beta z}+\epsilon_{\gamma\tau}(x_\gamma-x_{0\gamma})\overline\partial_\beta\omega^\star_\tau\nonumber,
\end{eqnarray}
from which, after multiplication by $\epsilon_{\kappa\alpha}$, it results that:
\begin{eqnarray}
&x_\alpha\overline\partial_\tau\omega_\kappa^\star=-\epsilon_{\kappa\alpha}\overline\partial_\tau b_z^\star
+\epsilon_{\kappa\alpha}\mathcal{E}_{\tau z}^\star+x_{0\alpha}\overline\partial_\tau\omega_\kappa^\star
+(x_\kappa-x_{0\kappa})\overline\partial_\tau\omega_\alpha^\star.\nonumber
\end{eqnarray}
Then, by lemma \ref{fleurs} with a permutation of $\kappa$ and $\alpha$, (\ref{ombre}) also writes as
\begin{eqnarray}\label{pik'}
\Pi'_\kappa=\partial_\alpha\varphi(\hat x)\int_{C_{\epsilon,z}}\left(-\epsilon_{\kappa \alpha}\overline\partial_\beta b_z^\star+\epsilon_{\kappa\alpha}\mathcal{E}_{\beta z}^\star + x_{0\alpha}\overline\partial_\beta\omega_\kappa^\star
-x_{0\kappa}\overline\partial_\beta\omega_\alpha^\star
\right)dx_\beta+o(1).\nonumber
\end{eqnarray}
On the other hand, from equation (\ref{ombre}) and lemma \ref{fleurs} (i.e. from strain compatibility) it follows that:
\begin{eqnarray}
\Pi'_\kappa&=&
\partial_\alpha\varphi(\hat x)\int_{C_{\epsilon,z}}\left(-\epsilon_{\kappa\beta}
\mathcal{E}_{\beta z}^\star dx_\alpha+\epsilon_{\alpha\beta}
\mathcal{E}_{\beta z}^\star dx_\kappa\right)+o(1)\nonumber\\&=&\partial_\alpha\varphi(\hat x)\int_{C_{\epsilon,z}}\epsilon_{\alpha\kappa}
\mathcal{E}_{\beta z}^\star dx_\beta+o(1).\label{chiho}
\end{eqnarray}
By summing this latter expression of $\Pi'_\kappa$ with  (\ref{chiho}), from the definitions of the Frank and Burgers vector it follows that
\begin{eqnarray}\label{bruno}
\Pi'_\kappa=\frac{1}{2}\partial_\alpha\varphi(\hat x)\epsilon_{\alpha\kappa}
\left(B_z^\star-\epsilon_{\gamma\beta}\Om_\gamma^\star x_{0\beta}\right)+o(1).
\end{eqnarray}
Hence, in the limit $\epsilon\to 0^+$  (\ref{bruno}) writes as
\begin{eqnarray}\label{fougeres}
\lim_{\epsilon\to 0^+}\Pi'_\kappa=\ll\left\{\frac{1}{2}\epsilon_{\kappa\alpha}B_z^{\star}
-\frac{1}{2}\epsilon_{\kappa\alpha}\epsilon_{\gamma\beta}\Om_\gamma^\star x_{0\beta}\right\}\partial_\alpha\delta_0,\varphi_z\gg.
\end{eqnarray}
$\blacktriangle$ Therefore, the result is proved on $\Om_z^0$, since
\begin{eqnarray}\label{buissons}
\lim_{\epsilon\to 0^+}\Pi_k(z,\varphi,\epsilon)=\lim_{\epsilon\to 0^+}\Pi'_k(z,\varphi,\epsilon)+\ll\Om^\star_k\delta_0,\varphi_z\gg.
\end{eqnarray}
As suggested by equation (\ref{FD1}), to obtain the result for the entire domain $\Om$ it suffices to integrate  equations (\ref{mousse}) and (\ref{bruno}) and the expression $\Om^\star_k\varphi(\hat x)$ over $L$, in order to replace $\delta_0$ by the line measure $\delta_L$ in (\ref{feuilles}), (\ref{fougeres}) and (\ref{buissons}). By (\ref{FD1}) the proof is then achieved.{\hfill $\square$}
\begin{theorem}[main result for a set of isolated defect lines]
 Let in the $2D$ case $L^{(i)},\ i\in\mathcal{I}\subset\NN$ stand for a set of isolated parallel dislocations and/or disclinations passing by $(\hat x_\beta^{(i)},z)$ and $\Om^{\star (i)}_z,B^{\star (i)}_k$ and $\delta_{L^{(i)}}$ denote the associated Frank and Burgers vectors, and the concentrated $1D$ Hausdorff measure density on $L^{(i)}$. Then under assumptions \ref{asstrain} and \ref{asstrainloc} in the vicinity of each defect line, incompatibility develops as the distribution 
\begin{eqnarray}
\eta_k^{\star}=\delta_{kz}\eta_z^{\star}+\delta_{k\kappa}\eta_\kappa^{\star}, \label{labo}
\end{eqnarray} 
with
\begin{eqnarray}
\eta^\star_z&=&\sum_{i\in\mathcal{I}}\left(\Om^{\star (i)}_z\delta_{L^{(i)}}+\epsilon_{\alpha\gamma}\left(B_\gamma^{\star (i)}+\epsilon_{\beta\gamma}(\hat x^{(i)}_\beta-x_{0\beta})\Om^{\star (i)}_z\right)\partial_\alpha\delta_{L^{(i)}}\right),\label{final1}\\
\eta^\star_\kappa&=&\frac{1}{2}\epsilon_{\kappa\alpha}\sum_{i\in\mathcal{I}}B^{\star (i)}_z\partial_\alpha\delta_{L^{(i)}}.\label{final2}
\end{eqnarray} 
\end{theorem}
{\bf Proof.} the proof is straightforward from theorem \ref{mainresultloc}. An alternative formulation is provided by (\ref{KR1})-(\ref{KR3}). {\hfill $\square$}

\subsection{Applications of the main result}\label{appl}
Throughout this section, $(x,y,z)$ denotes a generic point of $\Om_L$ and all tensors are written in matrix form in the Cartesian base $(e_x,e_y,e_z)$.
$\nl\bullet\ $\textit{Screw disclocation.} Since $B^\star_\gamma=\Om^\star_z=0$, (\ref{etaz}) and (\ref{etakappa}) yield
\begin{eqnarray}
\displaystyle[\eta^\star_k]=\frac{B^\star_z}{2}\left[\begin{array}{ccc} \partial_y\delta_L \\-\partial_x\delta_L \\ 0\end{array}\right].\nonumber
\end{eqnarray}
This result is easily verified with use of equation (\ref{FP2}). One needs to compute $<\eta^\star_k,\varphi>=\displaystyle \int_\Om\epsilon_{kpn}\epsilon_{\alpha\beta}\mathcal{E}^\star_{\beta n}\partial_p\partial_\alpha\varphi dV$,
that is to calculate the integral of
\begin{eqnarray}
\frac{B^\star_z}{4\pi}\left[\begin{array}{ccc} \partial_y\partial_x\varphi\frac{\cos\theta}{r}+\partial^2_y\varphi\frac{\sin\theta}{r}
\\  -\partial^2_x\varphi\frac{\cos\theta}{r}-\partial_x\partial_y\varphi\frac{\sin\theta}{r}
\\0\end{array} \right].\nonumber
\end{eqnarray}
By integration by parts, using Gauss-Green's theorem on $\Om$, and recalling that test-functions have compact supports and that $\displaystyle\partial_m\log r=\frac{x_m}{r^2}$, this integral becomes
\begin{eqnarray}
-\frac{B^\star_z}{4\pi}\displaystyle\int_\Om
\left[
\begin{array}{c}
\partial_y\varphi\left(\partial_x\frac{\cos\theta}{r}
+\partial_y\frac{\sin\theta}{r} \right)\\ -\partial_x\varphi\left(\partial_x\frac{\cos\theta}{r}+\partial_y\frac{\sin\theta}{r}\right)\\
0
\end{array}
\right]dV
=
\frac{B^\star_z}{4\pi}\displaystyle\int_\Om
\left[
\begin{array}{c} -\partial_y\varphi\partial_m^2\log r\\
\partial_x\varphi\partial_m^2\log r\\
0
\end{array}
\right]dV.\nonumber
\end{eqnarray}
Hence, from the relation $\Delta\left(\log r\right)=2\pi\delta_L$, the first statement is  verified.

$\nl\bullet\ $\textit{Edge dislocation.} Whereas $\overline\partial_m\omega^\star_k$ identically vanishes on $\Om_L$,  it is easily seen that (\ref{etaz}) and (\ref{etakappa}) with $B^\star_z=\Om^\star_z=0$ yield
\begin{eqnarray}
[\eta^\star_k]=B^\star_y\left[\begin{array}{ccc} 0\\0 \\ \partial_x\delta_L
\end{array}\right].\nonumber
\end{eqnarray}
We must compute $\displaystyle <\eta^\star_k,\varphi>=\int_\Om\epsilon_{kpn}\epsilon_{\alpha\beta}\mathcal{E}^\star_{\beta n}\partial_p\partial_\alpha\varphi dV$.
For $n\not =3$, the strain components do not identically vanish and, for $k=1$ and $k=2$, we must have $p=3$ and hence the only non-vanishing component of the expression $\epsilon_{\alpha\beta}\mathcal{E}^\star_{\beta n}\partial_p\partial_\alpha\varphi$ are $\mathcal{E}^{\star}_{yx}\partial_z\partial_y\varphi-\mathcal{E}^{\star}_{yy}\partial_z\partial_x\varphi$ and
$\mathcal{E}^{\star}_{xy}\partial_z\partial_x\varphi-\mathcal{E}^{\star}_{xx}\partial_z\partial_y\varphi$. By integration by parts, recalling that the strain does not depend on $z$, the related integrals vanish. For $k=3$, the integrand is
\begin{eqnarray}
\epsilon_{pnz}\epsilon_{\alpha\beta}\mathcal{E}^\star_{\beta n}\partial_p\partial_\alpha\varphi=
\left(\partial_y\mathcal{E}^\star_{xx}-\partial_x\mathcal{E}^\star_{xy}\right)\partial_y\varphi
+\left(\partial_y\mathcal{E}^\star_{xy}-\partial_x\mathcal{E}^\star_{yy}\right)\partial_x\varphi
.\nonumber
\end{eqnarray}
By inserting the expression of the strain tensor into the right-hand side of this equation, integration by parts provides the expression $\displaystyle\int_\Om-\frac{B_y}{2\pi}\partial_x\varphi \Delta(\log r)dV$, achieving the second verification. 
\newline
$\nl\bullet\ $\textit{Wedge disclination.} Incompatibility reads
\begin{eqnarray}
[\eta^\star_k]=\Om^\star_z\left[\begin{array}{ccc} 0\\0 \\ \delta_L\end{array}\right].\nonumber
\end{eqnarray}
We must calculate $<\eta_k^\star,\varphi>=\displaystyle\int_\Om\epsilon_{kpn}\epsilon_ {\alpha\beta}\strain_{\beta n}\partial_p\partial_\alpha\varphi dV$. For $k=1$ and $k=2$, we must have $n\not =3$ and $p=3$, but then the integrand vanishes. For $k=3$, we compute
\begin{eqnarray}
\epsilon_{pn}\epsilon_{lm}\mathcal{E}^{\star}_{mn}\partial_p\partial_l\varphi&=&
\frac{\Om^\star_z(1-\nu^\star)}{4\pi}\varphi\Delta(\log\frac{r}{R})
+\frac{\Om^\star_{z}(1+\nu^\star)}{4\pi}\varphi\Delta(\log\frac{r}{R})\nonumber\\
&=&2\frac{\Om^\star_z}{4\pi}\varphi(2\pi\delta_L),\nonumber
\end{eqnarray}
achieving the third verification.
\section{Conclusive remarks}\label{concl}
In this paper a general theory revisiting the work of Kr\"{o}ner (1980) has been developed to model line defects in single crystals at the mesoscopic scale. A rigorous definition of the dislocation and disclination density tensors as concentrated effects on the defect lines has been provided in the framework of the distribution theory. The main difficulty resulting from the multivaluedness of the displacement and rotation vector fields in defective crystals has been addressed by defining the single-valued Burgers and Frank tensors from the distributional strain gradient. Whereas outside the defective lines both tensors are regular functions directly related to the displacement and rotation gradients, in addition they exhibit concentrated properties within the defect lines which may be linked to the displacement and rotation jumps around these lines.

Moreover, defining the incompatibility tensor as the distributional curl of the Frank tensor, the principal result of our work has been to express in the two-dimensional case incompatibility as a function of the dislocation and disclination density tensors and their distributional gradients, and to demonstrate this relationship under precise assumptions on the regularity of the strain tensor in the vicinity of the assumed isolated defect lines. In a subsequent paper (Van Goethem \& Dupret, 2009$b$), our theory is extended to the case of a countable number of defect lines under specific hypotheses based on the geometric measure theory.

In general our work is devoted to provide a rigorous distributional definition and a new understanding of the different mathematical objects (dislocation and disclination densities, contortion, incompatibility, Burgers and Frank tensors, elastic strain, etc.) that can be added at the mesoscopic scale in order to well-define the associated homogenized objects at the macroscopic scale. Further work will deal with the general three-dimensional dynamic theory.